\documentclass[%
 reprint,
nofootinbib,
 amsmath,amssymb,
 aps,
prb,
]{revtex4-1}
\usepackage{graphicx}% Include figure files
\usepackage{dcolumn}% Align table columns on decimal point
\usepackage{bm}% bold math
\usepackage{xcolor}
\usepackage{caption}
\usepackage{textcomp}
\makeatletter
\newcommand{\mathleft}{\@fleqntrue\@mathmargin0pt}
\newcommand{\mathcenter}{\@fleqnfalse}
\makeatother

\newcommand{\rev}[1]{ #1 }

\begin{document}

\preprint{APS/123-QED}

%referees: tolias,bostrom,bergstrom,kimbal,palasantzas

\title{Analytical theory for the crossover  from retarded \\ to non-retarded interactions between metal plates}%

\author{Juan Luengo-M\'arquez}
\email{juan.luengo@uam.es}
\affiliation{%
Department of Theoretical Condensed Matter Physics and Instituto Nicol\'as Cabrera, \\
Universidad Autónoma de Madrid, 28049 Madrid (Spain)
}%
\author{Luis G. MacDowell}%
\email{lgmac@quim.ucm.es}
\affiliation{%
Departamento de Qu\'imica F\'isica, Facultad de Ciencias Químicas, Universidad Complutense de Madrid, 28040, Madrid, Spain.
}%


%\affiliation{
%\affiliation{
%\affiliation{%


\date{\today}% It is always \today, today,

%\begin{abstract}  
%	The van der Waals force established between two surfaces %plays a central role in many phenomena, such as adhesion or %friction\cite{israelachvili91,parsegian05,munday09}.  %According to Lifshitz %theory\cite{dzyaloshinskii61,lifshitz56,parsegian05}, in the %low temperature limit the surface  interaction may exhibit %widely different regimes depending on the distance of %separation \cite{palasantzas08}. At short separation, results %are known for a broad variety of systems with great accuracy, %including the interaction between metallic %surfaces\cite{tolias18,tolias20}. The Casimir force - %characteristic of the long distances regime - has been also %largely explored for a system composed of two metal %plates\cite{bostrom04}. However, the crossover regime has %defied analytical approximations and is far less well %understood. In this study, we adopt a Quadrature - based %methodology\cite{macdowell19,luengo21,fiedler21} to provide %new insight into the physical features behind the interplay %between two plates of metal across vacuum at finite %temperature. Moreover, we present a novel methodology that %exploits %the mean value theorem to compute the Hamaker %constant accurately in analytical form. We also display a %fully-analytic %solution of the retarded interactions based on %the Drude model, comparing the outcome with the exact 
%Lifshitz result.
%\end{abstract}

\begin{abstract}  
	The van der Waals force established between two surfaces plays a central role 
	in many phenomena, such as adhesion or friction. However, the dependence of this 
	forces on the distance of separation between plates is very complex. Two widely
	different non-retarded and retarded regimes are well known, but these have
	been traditionally studied separately. Much less is known about the
	important experimentally accessible cross-over regime.  In this study, we
	provide analytical approximations for the van der Waals forces between two
	plates that interpolates exactly between the short distance and long
	distance behavior, and provides new insight into the crossover from London
	to Casimir forces at finite temperature. At short distance, where the
	behavior is dominated by non-retarded interactions, we work out a
	very accurate simplified approximation for the Hamaker constant which
	adopts analytical form for both the Drude and Lorentz models of dielectric
	response. We apply our analytical expressions for the study of forces
	between metallic plates, and observe very good agreement with exact
	results from numerical calculations. Our results show that contributions
	of interband transitions remain important in the experimentally accesible
	regime of decades nm for several metals, including gold.
\end{abstract}

\maketitle

%\begin{figure}[h!]
%\includegraphics[scale=0.65]{fig_abstract.pdf}
%\end{figure}

%\begin{itemize}
%\item The adsorption equilibrium on an inert substrate at three phase
%coexistence is formulated and the conditions for wetting of either of
%the three phases is formulated in terms of a generalized surface potential.
%\item The explicit expressions for Dzyaloshinskii-Lifshitz-Pitaevskii
%for a system of two variable adsorption layers is worked out.
%\item Analytic expressions for the van der Waals forces are provided
%which allow to calculate surface forces from the Hamaker non-retarded to
%the Casimir retarded regimes with the same data that 
%is conventionally used to calculate Hamaker constants.
%\item We show that at the three phase contact line between AgI/water/air,
%van der Waals forces promote growth of ice both on the AgI/air and the
%water/vapor interfaces, lending support to a contact mode nucleation
%mechanism of AgI in the atmosphere.
%\end{itemize} 

\section{Introduction}

Dispersion forces arise between any two polarizable media, and are therefore
ubiquitous in nature.  Even if their strength is often  relatively weak at short
range\cite{derjaguin87}, they appear to have a significant contribution to the
explanation of a wide range of phenomena
\cite{israelachvili91,parsegian05,ninham10}, such as ice premelting
\cite{fiedler20,esteso20}, the stabilization of thin lipid
films\cite{parsegian69}, or the adhesion of Geckos to vertical surfaces\cite{autumn00}. 
\vspace{0.1cm}
\\
The characterization of  interactions between macroscopic bodies is often
described as the result of the summation over all individual interactions
between pairs of particles\cite{hamaker37,dietrich88,schick90,israelachvili91}. For the interaction between two plates across vacuum, this view results in an energy function, $g(h)$, that decreases as the squared inverse distance of separation between the plates, $h$:
\begin{equation}
	g(h) = -\frac{A_{ham}(0)}{12\pi h^{2}}
	\label{eq:energy}
\end{equation}
The Hamaker constant, $A_{ham}(0)$ is a fundamental  property of the interacting bodies that lumps the magnitude of the force established between pairs of molecules. From this perspective, it is a mean field parameter that is dictated by properties of pairs of interacting particles  within two macroscopic objects. 
\vspace{0.1cm}
\\
Alternatively, the more modern Lifshitz theory of the van der Waals forces\cite{dzyaloshinskii61,lifshitz56} computes the Hamaker constant considering instead the energies assigned to the electromagnetic modes of vibration allowed inside the system. The dispersion forces emerge in this approach from simultaneous fluctuations of the particles  as a response to these electromagnetic waves. In this framework, $A_{ham}(0)$ is dictated instead by the collective dielectric response of the materials.

However, when the distance of separation takes large values, the picture is much more complex. As $h$ increases, the electromagnetic waves require significant amounts of time to promote fluctuations between material patches. This happens due to the fact that the speed of light at which the electromagnetic waves move is finite, so that when the vibration frequency of the polarizable particles is large, the  time of the wave to travel from atom to atom might be comparable to the period assigned to the fluctuations. This phenomenon is called retardation\cite{parsegian05}, and effectively weakens the dispersive interactions for large distances.
\vspace{0.1cm}
\\
The retardation effect is considered also in the Lifshitz theory of van der Waals forces \cite{lifshitz56,dzyaloshinskii61}. When taken into account, the Hamaker coefficients of Eq. \ref{eq:energy} becomes a function of the distance of separation, $A_{ham} = A_{ham}(h)$. For distances close to zero, $A_{ham}(h\rightarrow 0)$ reaches a constant value expected in the absence of retardation, the Hamaker constant $A_{ham}(0)$. This short-distances regime is dubbed London regime, i.e., we talk about London dispersion forces.
\vspace{0.1cm}
\\

On the other hand, as the distance of separation between the plates increases, $A_{ham}(h)$ decays gradually and develops a distinct $h$ dependence, which cannot be described in terms of pairwise summation of dispersion interactions. Particularly, for two perfect metal plates in vacuum at large separation and zero temperature, Casimir showed  that the energy of interaction adopts the celebrated form \cite{casimir48}:
\begin{equation}\label{eq:casimir}
  g(h) = -\frac{\pi^2 c\hbar}{720 h^3}
\end{equation}
This result explicitly points to the non-trivial dependence of the 'Hamaker
constant', which, more accurately corresponds to a Hamaker function $A(h)$
decaying as $1/h$ in this limit
\cite{israelachvili72,sabisky73,white76,chan77,gregory81,palasantzas08,vanzwol10}. 

As a matter of fact, the interest of this result goes well beyond the mere study of surface interactions, as it conveys invaluable information on the nontrivial structure of the quantum vacuum and the role of zero-point energy in physics \cite{milton98,lamoreaux07}. For this reason, there has been considerable interest in verifying this prediction experimentally \cite{lamoreaux97,mohideen98,bressi02,munday09,man09,garret18}. In practice, however, it must be recognized that Eq.\ref{eq:casimir} is an asymptotic result which can only be realized at zero temperature, for perfect metals. Verification of the underlying physics of Eq.\ref{eq:casimir} needs to take into account the simultaneous effect of finite temperature and finite conductivity of metals \cite{obrecht07,fisher20}. Nevertheless, attempts to single out asymptotic corrections to the more general result of Lifshitz and collaborators \cite{lifshitz56,dzyaloshinskii61} is difficult, and remains
still a matter of discussion
\cite{ninham70,parsegian70,chan77,schwinger78,milton98,ninham98,bostrom00,lambrecht00,bezerra04,geyer05,ninham14,obrecht07,fisher20}. 

Recently, a promising approach for the study of the crossover regime from retarded to non-retarded interactions was proposed \cite{macdowell19,luengo21}. The result provided well known exact analytic results over the full range of plate separations, albeit within the so called dipole approximation of the Lifshitz equation. The key idea in that study is a quadrature rule which allows to describe the crossover regime in a non-perturbative manner. This is a significant issue, because in the Lifshitz result, plate separation, temperature and dielectric properties are entangled in a highly non-trivial manner, so it is not clear whether each effect can be singled out separately.

In this study, we aim to extend that work beyond the linear dipole  approximation, which is required in order to obtain the exact Casimir limit for perfect metals at zero temperature. 

In the next section we provide the essential background to the Lifshitz theory of intermolecular forces.
 In section III we extend the recently introduced Weighted Quadrature
 Approximation (WQA) beyond the dipole approximation. The working formulae
 relies on knowledge of the exact Hamaker constant at zero plate separation.
 Therefore, we devote section IV to derive a new, simple and accurate
 approximation for the Hamaker constant $A_{ham}(0)$. In the next section, we
 work out analytical results for retarded interactions between materials obeying
 the Drude model of the dielectric responses. 
 In Section VI
 we compare the resulting interaction coefficients with exact numerical
 solutions of Lifshitz theory calculated from a detailed description of 
 dielectric properties published recently \cite{tolias18,gudarzi21}, 
 showing how the new
 methodology for the computation of the Hamaker constant proposed in this
 article yields values of $A_{ham}(0)$ in excellent agreement with the ones
 predicted by the exact Lifshitz formula. Our results are summarized in the
 Conclusions.

\section{Lifshitz theory}

In the frame of Lifshitz theory, the Hamaker function for retarded interactions
between two metal plates across vacuum takes the form\cite{supplementary21}
	\mathleft
\begin{equation}	
 A_{ham}(h) = 
\notag
\end{equation}
\mathcenter
\begin{equation}
	\frac{3 k_{B}T}{2}{\sum_{n=0}^{\infty}}'\int_{r_{n}}^{\infty}\sum_{k=1}^{\infty}x \ dx \frac{(\Delta^{E}_{mv})^{2k}+(\Delta^{M}_{mv})^{2k}}{k}e^{-kx}
\label{eq:lifshitz}
\end{equation}
Where $r_{n} = 2h\epsilon_{v}^{1/2}\omega_{n}/c$,  $k_{B}$ is the Boltzmann's constant and $T$ is the temperature in Kelvin. All along this study we employ room temperature, $T=300\ K$. Besides, we have that
\begin{equation}
	\Delta^{E}_{mv} = \frac{x_{v}-x_{m}}{x_{v}+x_{m}},\hspace{1 cm} \Delta^{M}_{mv} = \frac{\epsilon_{m}x_{v}-\epsilon_{v}x_{m}}{\epsilon_{m}x_{v}+\epsilon_{v}x_{m}}
	\label{eq:delta}
\end{equation}
With $x_{i}^{2} = x^{2}+(\epsilon_{i}-\epsilon_{v})(2h\omega_{n}/c)^{2}$, being
$c$ the speed of light. The subscripts $m$ and $v$ denote metal and vacuum,
respectively. In these equations, the magnetic susceptibilities of the media
have been assumed to be equal to unity. The $\epsilon_{i}$ is the dielectric response of each medium $i$, which reflects the tendency of that substance to polarize reacting to an electromagnetic wave. It is thus a function of the frequency of the incoming wave, which is also the frequency at which the particles of that material will oscillate in response\cite{parsegian05}. Here the $\epsilon_{i}(\omega_{n})$ are evaluated at the discrete set of Matsubara frequencies $\omega_{n} = 2\pi n k_{B}T/\hbar$, being $\hbar$ the Planck's constant in units of angular frequency.
\vspace{0.1cm}
\\
Notice that the result of \ref{eq:lifshitz} greatly generalizes the well known
result of Casimir, and allows for a  panoply of interaction, including
non-monotonous dependence of $A_{ham}(h)$, as well as monotonic repulsive interactions. For instance, retardation-driven repulsion has been found for the force between gold and silica surfaces immersed in bromobenzene\cite{munday09,bostrom12}. This repulsive interaction in the Casimir regime has been found to allow supersliding between two surfaces, arising from an extremely low friction coefficient\cite{feiler08}.
\vspace{0.1cm}

The prime in Eq. \ref{eq:lifshitz} means that the first term of the summation in
$n$ has half weight. This term corresponds to the contribution of the zero
frequency, the only one that remains as $h\rightarrow \infty$, and it thus
provides the interaction coefficient for very large distances. Singling out
this term, the zero frequency contribution is given as\cite{supplementary21}:
\begin{equation}
	A_{ham}^{\omega_{n}=0} = \frac{3k_{B}T}{4}\sum_{k=1}^{\infty}\left(\frac{\epsilon_{m}(0)-\epsilon_{v}(0)}{\epsilon_{m}(0)+\epsilon_{v}(0)}\right)^{2k}\frac{1}{k^{3}}
	\label{eq:zero}
\end{equation}
Where $\epsilon_{i}(0)$ is the static dielectric response, which for metals goes to infinity. Consequently, Eq. \ref{eq:zero} reveals that the interaction of two plates of metal across vacuum has  $A_{ham}^{\omega_{n}=0} = A_{ham}(h\rightarrow \infty) = \frac{3k_{B}T}{4} \zeta(3)$, which, at $T=300$~K amounts to $3.73\times 10^{-21} \ J$.
\\
After these considerations, we can focus on the remaining contributions of Eq.
\ref{eq:lifshitz}, namely $A_{ham}^{\omega_{n}>0}(h)$. A usual treatment of the
Hamaker function is to consider only the first term of the summation in $k$
\cite{tabor69,hough80,prieve88,bergstrom97}. This is called the linear or dipole approximation, and works well in those cases where the $ (\Delta^{E}_{mv})^{2k}+(\Delta^{M}_{mv})^{2k}$ function vanishes rapidly. In the situation that we handle, the large values of $\epsilon_{m}(\omega_{n})$, specially for low frequencies due to the plasma resonance, make the use of the linear approximation unreliable.
\vspace{0.1cm}
\\
Additionally, the complex interplay between London and Casimir regime that we have described is encapsulated inside the Eq. \ref{eq:lifshitz} in a non trivial way. In previous works\cite{macdowell19,luengo21} we have taken advantage of several mathematical tools to provide insightful expressions within the linear approximation attempting to clarify the physical interpretation of the Lifshitz formula. 
\vspace{0.1cm}
\\

\section{WQA beyond the dipole approximation}

The Weighted Quadrature Approximation (WQA) introduced recently  within the
linear dipole approximation\cite{macdowell19}, employs the Gaussian Quadrature
as an analytical tool to simplify the Lifshitz formula. Here we use the same idea to generalize the WQA to the infinite
order sum of  Eq. \ref{eq:lifshitz}. After a first Gaussian Quadrature, using
$xe^{-kx}$ as the weight function, and the approximation of the summation in $n$
to an integral via the Euler-MacLaurin formula, we reach\cite{supplementary21}
\begin{equation}
A_{ham}^{\omega_{n}>0}(h) = \frac{3c\hbar}{8\pi}\int_{\nu_{T}}^{\infty}d\nu \ \sum_{k = 1}^{\infty} \widetilde{R}_{k}(\nu,x_{1,k},h)[k\nu h+1]e^{-k\nu h}
	\label{eq:start}
\end{equation}
Where $\nu_{T} = 2\epsilon_{v}^{1/2}\omega_{T}/c$, $\omega_{T} = 2\pi k_{B}T/\hbar$  and $\widetilde{R}_{k}(\nu,x_{1,k},h) = \epsilon_{v}^{-1/2}j_{v}^{-1}R_{k}(\nu,x_{1,k},h)$, being $j_{v} = \left(1+\frac{1}{2}\frac{d \ln \epsilon_{v}}{d \ln \omega_{n}}\right)$, $x_{1,k} = (kr_{n}^{2} + 2r_{n} + 2/k)/(kr_{n} + 1)$, and 
\begin{equation}
	R_{k}(\nu,x,h) = \frac{(\Delta^{E}_{mv})^{2k}+(\Delta^{M}_{mv})^{2k}}{k^{3}}
	\label{eq:erre}
\end{equation}
In practice, for metallic plates interacting across vacuum,  $\epsilon_{v}(\omega)=1$, so that $j_{v}=1$, and  $\widetilde{R}_{k}(\nu,x,h)= R_{k}(\nu,x,h)$.

Recall that the dependence on $\nu=\nu_{T}n$, $x$, and $h$ enters by the hand of $\Delta^{E,M}_{mv}$, as dictated by Eq. \ref{eq:delta}. The function $R_{k}(\nu,x,h)$ exhibits a very complicated dependence on the distance of separation, arising from the fact that the frequencies captured by the integral in Eq. \ref{eq:start} are being reduced as the distance increases. At this stage, this result is still too arid to infer intuitively the expected functional behavior. 
\vspace{0.1cm}
\\
We proceed by introducing $\nu_{\infty}$, an effective parameter specific of
each system  meant to describe the decay length of $R_{k}(\nu,x,h)$ as
$\nu\to\infty$. Then a second Gaussian Quadrature is
performed\cite{supplementary21}, and we achieve the WQA extended to include the complete summation over k
\begin{equation}
A_{ham}^{\omega_{n} >0}(h) = \frac{3c\hbar\nu_{\infty}}{8\pi}\sum_{k=1}^{\infty}\widetilde{R}_{k}(\nu^{*}_{k},x_{1,k},h)e^{\xi_{k}}e^{-\nu_{T}kh}\widetilde{F}_{k}
\label{eq:wqa}
\end{equation}
\begin{equation}
\widetilde{F}_{k} = \frac{(\nu_{T}kh+1)(\nu_{\infty}kh+1)+\nu_{\infty}kh}{(\nu_{\infty}kh+1)^{2}}
\notag
\end{equation}
\begin{equation}
 \xi_{k}= \frac{(\nu_{T}kh+1)(\nu_{\infty}kh+1)+2\nu_{\infty}kh}{(\nu_{\infty}kh+1)^{2}(\nu_{T}kh+1)+(\nu_{\infty}kh+1)\nu_{\infty}kh}
\notag
\end{equation}
With  $\xi_{k} = (\nu^{*}_{k} - \nu_{T})/\nu_{\infty}$.

Truncating the sum in Eq.\ref{eq:wqa} beyond $k=1$, this result becomes the original WQA proposed recently\cite{macdowell19}. To this order of approximation, WQA is very accurate for dielectric materials with low dielectric response\cite{macdowell19,luengo21}. However, the extension provided here is required to describe interactions for materials with large dielectric response, because the terms $\Delta^{M/E}_{mv}$ are close to unity and the convergence of the series is not fast enough to warrant truncation at first order.

Equation \ref{eq:wqa} has not only the advantage of being entirely analytic, but also allows straightforward interpretation of the transition from London to Casimir regime of $A_{ham}(h)$ through the comparison of the magnitudes of $h$, $\nu_{\infty}$ and $\nu_{T}$.
\vspace{0.1cm}
\\
For short distances, $h\ll \nu_{\infty}^{-1}$, 
\rev{we find the auxiliary functions $\widetilde{F}_k=1$ and $\xi_k=1$, so that
	\begin{equation}
	A_{ham}^{\omega_{n} >0}(h\to 0) = \frac{3c\hbar\nu_{\infty}}{8\pi}\sum_{k=1}^{\infty}
	\left( \frac{\epsilon_m-1}{\epsilon_m+1}  \right)^{2k} \frac{e}{k^3}
	\label{eq:lim0}
	\end{equation}
with the dielectric function evaluated at a constant wave number $\nu^{*}=\nu_T + \nu_{\infty}$. Accordingly, the value of $A_{ham}(h\rightarrow 0)$ in Eq. \ref{eq:wqa} becomes independent of $h$, and the usual Hamaker constant is recovered.
}
\rev{
At large values of the distance of separation $h\gg \nu_{T}^{-1}$, 
we find $\xi_k\to 0$, while $\widetilde{F}_k=\nu_T/\nu_{\infty}$. This yields:
\begin{equation}
 A_{ham}^{\omega_{n} >0}(h\to\infty) = \frac{3c\hbar\nu_{T}}{4\pi}\sum_{k=1}^{\infty} \left( \frac{\epsilon_m^{1/2}-1}{\epsilon_m^{1/2}+1}\right)^{2k} \frac{e^{-\nu_{T}kh}}{k^3}
 \label{eq:liminf} 
\end{equation}
where the dielectric functions are now evaluated at the thermal wave-number $\nu_T$. In this limit, the retarded interactions are supressed exponentially. Accordingly, only the static term in the Hamaker constant survives.  This leads to the other asymptotic behavior $A_{ham}(h\rightarrow \infty)$, also independent of $h$.
}
\rev{
 In between these two limits, $\nu_{\infty}^{-1} \ll $h$ \ll \nu_{T}^{-1}$, the
 $h$ dependence of $A_{ham}(h)$ is governed by the factor
 $\widetilde{F}_{k}(h)$, which in this range  effectively decays as
 $2/(\nu_{\infty}kh)$. This leads to:
 \begin{equation}
 A_{ham}^{\omega_{n} >0}(h) = \frac{3c\hbar}{4\pi h}\sum_{k=1}^{\infty} \frac{\left( \Delta_{m}^E +  \right)^{2k} + \left(\Delta_m^M \right)^{2k}}{k^4}
 \label{eq:limh} 
 \end{equation}
 where the dielectric functions are now to be evaluated at frequencies lying between $\nu_T$ and $\nu_{\infty}$.
 For a perfect metal, implying $\Delta_{mv}^{E/M}=1$, Eq.\ref{eq:wqa} readily yields 
$A_{ham}^{\omega_{n} >0}(h) = \frac{3c\hbar}{2\pi h}\zeta(4)$, which corresponds to the exact result of Casimir for two interacting metals at zero temperature.
}
\vspace{0.1cm}
\\
This inspection highlights one major strength of the WQA: it provides a parametric clarification of the qualitative change of the distance dependence on the dispersion forces. The distance implied in $\nu_{\infty}^{-1}$ signals the point at which the surface van der Waals free energy in Eq. \ref{eq:energy} switches its dependence on the distance of separation from the $\sim 1/h^{2}$ proper of London regime, to the $~1/h^{3}$ characteristic of Casimir interactions which is exact at zero temperature. At finite temperature, however, $\nu_T$ is finite, and the Casimir regime becomes fully suppressed for $h>\nu_T^{-1}$.

\section{A simple Quadrature method for the Hamaker constant}

For practical matters, use of Eq. \ref{eq:wqa} requires knowledge of the system's parameter $\nu_{\infty}$, \rev{which must be chosen so as to obtain optimal agreement with the exact results. A look at the limiting regimes of $A_{ham}^{\omega_{n}>0}(h)$ displayed in Eq.\ref{eq:lim0}-\ref{eq:limh} shows that  the choice of $\nu_{\infty}$ is only significant in the limit of small $h$. Accordingly, we seek $\nu_{\infty}$ by matching  the $h\rightarrow 0$ limit of Eq. \ref{eq:wqa} to the exact  $A_{ham}^{\omega_{n}>0}(0)$ from Eq. \ref{eq:lifshitz} (c.f. Ref.\cite{luengo21}).}  Unfortunately, the numerical calculation of $A_{ham}^{\omega_{n}>0}(0)$ also is a cumbersome task. Therefore, in this section we present a very accurate one-point quadrature for $A_{ham}(0)$ that adopts a particularly simple analytic form for two metallic plates obeying the Drude model. 

Our starting point is   Eq.\ref{eq:start} in the limit of $h\rightarrow 0$: 
\begin{equation}
A_{ham}^{\omega_{n}>0}(h\to 0) = \frac{3c\hbar}{8\pi}\int_{\nu_{T}}^{\infty}d\nu \ Li_3(R_1)
\label{eq:start0}
\end{equation}
where $Li_3(x)=\sum_{k=1} x^k/k^3$ is the polylogarithmic function, and
$R_1(\nu)=R_1(\nu,x,h=0)$ takes the particularly simple form:
\begin{equation}
R_{1}(\nu) = \left(\frac{\epsilon_{m}(i\nu)-\epsilon_{v}(i\nu)}{\epsilon_{m}(i\nu)+\epsilon_{v}(i\nu)}\right)^{2}
\label{eq:erre1}
\end{equation}
%$R_1(\nu)=[(\epsilon_m-\epsilon_v)/(\epsilon_m+\%epsilon_v)]^2$. 
 Analytical approximations for the Hamaker constant are usually obtained by replacing $Li_3(R_1)$ by its zeroth order approximation $Li_3(R_1)\approx R_1$ and truncating beyond first order.  Here, we notice that $R_1(\nu)$ has the properties of a well behaved distribution, so that applying the first mean value theorem, Eq.\ref{eq:start0} may be expressed exactly as:
\begin{equation}
A_{ham}^{\omega_{n}>0}(h\to 0) =
\frac{3c\hbar}{8\pi} 
 \frac{Li_3(R_1^*)}{R_1^*} 
\int_{\nu_{T}}^{\infty} d\nu {R}_{1}(\nu)
\label{eq:start1}
\end{equation}
where $R_1^* = R_1(\nu*)$, is evaluated at a mean value frequency, $\nu*$, in
the interval between $\nu_T$ and $\infty$. An accurate estimate of $R^*_1$,
without needing an explicit evaluation of $\nu*$ may be obtained right away by
requiring the second order expansion of the quadrature rule Eq.\ref{eq:start1}
to match exactly that of  Eq.\ref{eq:start0}. This then leads to the simplified
prescription (see supplementary material, Ref.\cite{supplementary21}):
\begin{equation}
A_{ham}^{\omega_{n}>0}(h\to 0) =
\frac{Li_3(R^*_1)}{R^*_1} \,
A_{ham,0}^{\omega_{n}>0}  
\label{eq:quadrature}
\end{equation}
where
\begin{equation}
R^{*}_1 = \frac{I_{2}}{I_{1}}, \hspace{1cm} I_{m} = \int_{\nu_{T}}^{\infty}\left[R_{1}(\nu)\right]^{m}d\nu
\notag
\end{equation}
while $A_{ham,0}^{\omega_{n}>0}=\frac{3c\hbar}{8\pi}I_1$ is just the usual zeroth order approximation of the Hamaker constant.

In practice, $R_1(\nu)<1$  at all frequencies, so that $R_1^*$ is small always
and $Li_3(R_1^*)/R^*_1=1+R_1^*/8+(R_1^*)^2/27...$ is a  correction factor of order unity.
In fact, by construction, Eq.\ref{eq:quadrature} is exact up to second order in
$R_1^*$, and provides very accurate results when compared with the numerical
solution of Eq. \ref{eq:lifshitz}. Henceforth, we will refer to the prescription
of Eq.\ref{eq:quadrature} as the Q rule. Its performance will be tested later
for the interaction between two plates of Al, Be, Cr, W and Au using 
recently published data as a benchmark.\cite{tolias18,gudarzi21}

\begin{figure*}[htb!]
	\includegraphics[width=0.48\textwidth,height=0.35\textwidth,keepaspectratio]{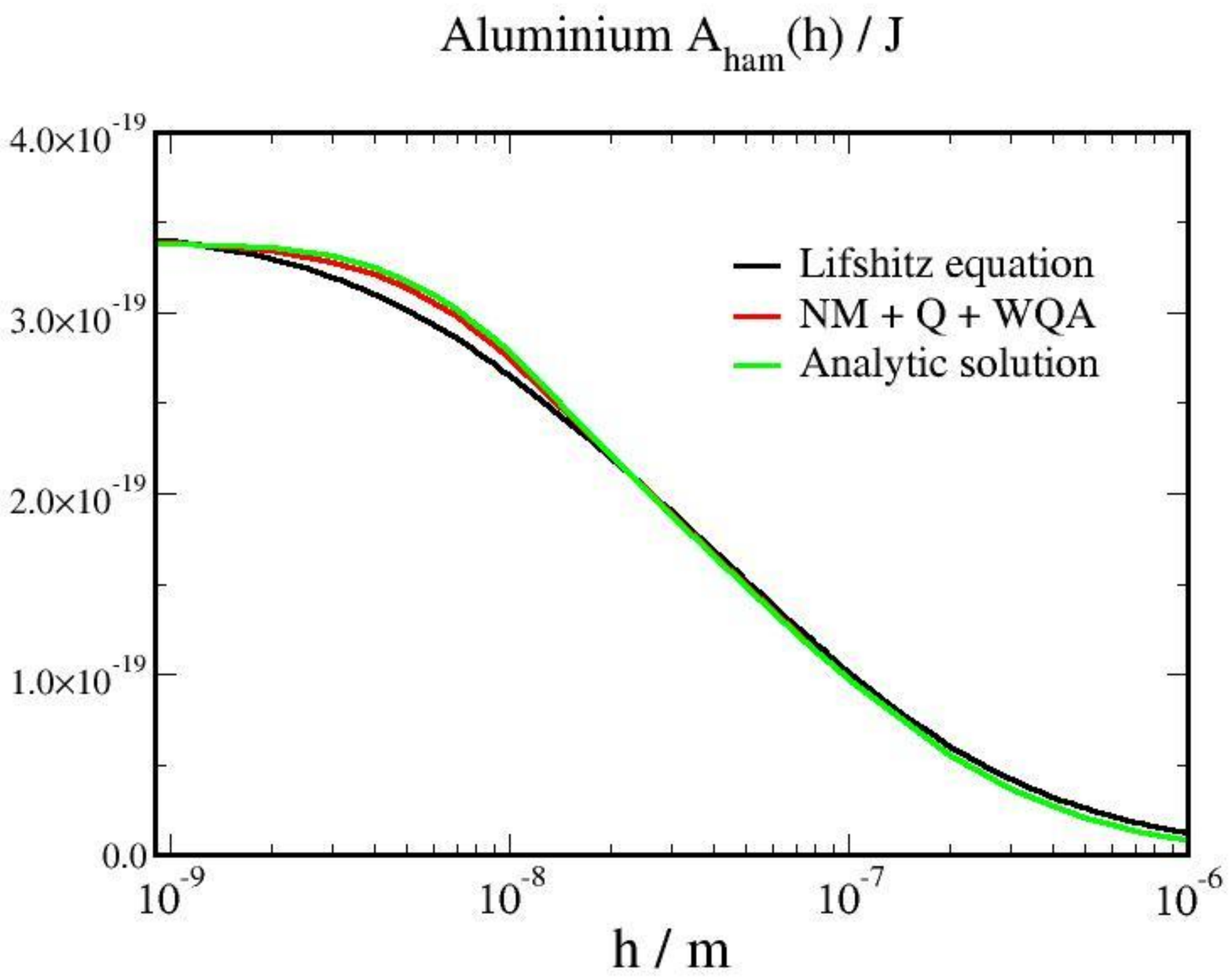}
	\includegraphics[width=0.48\textwidth,height=0.35\textwidth,keepaspectratio]{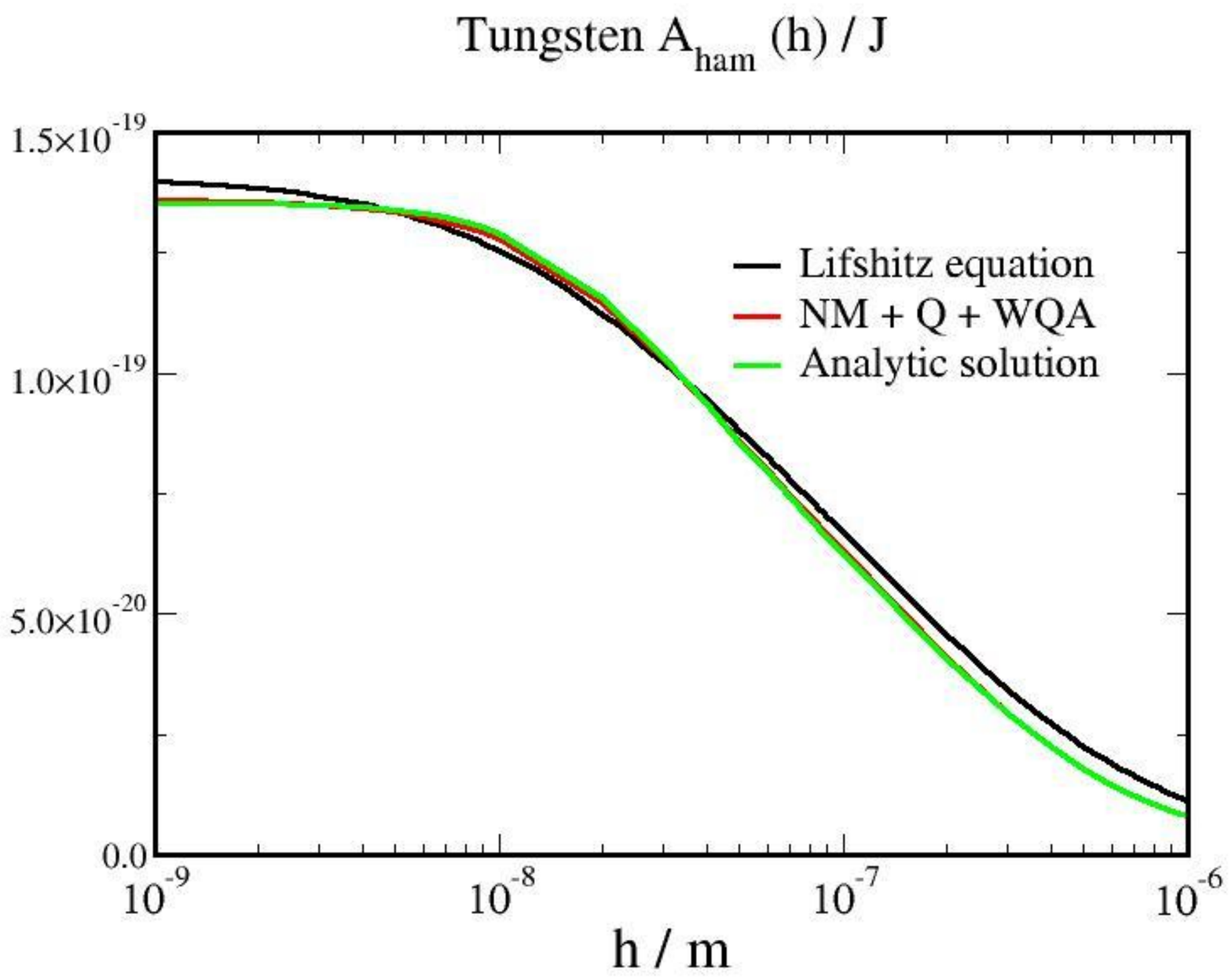}
	\\
	\includegraphics[width=0.48\textwidth,height=0.35\textwidth,keepaspectratio]{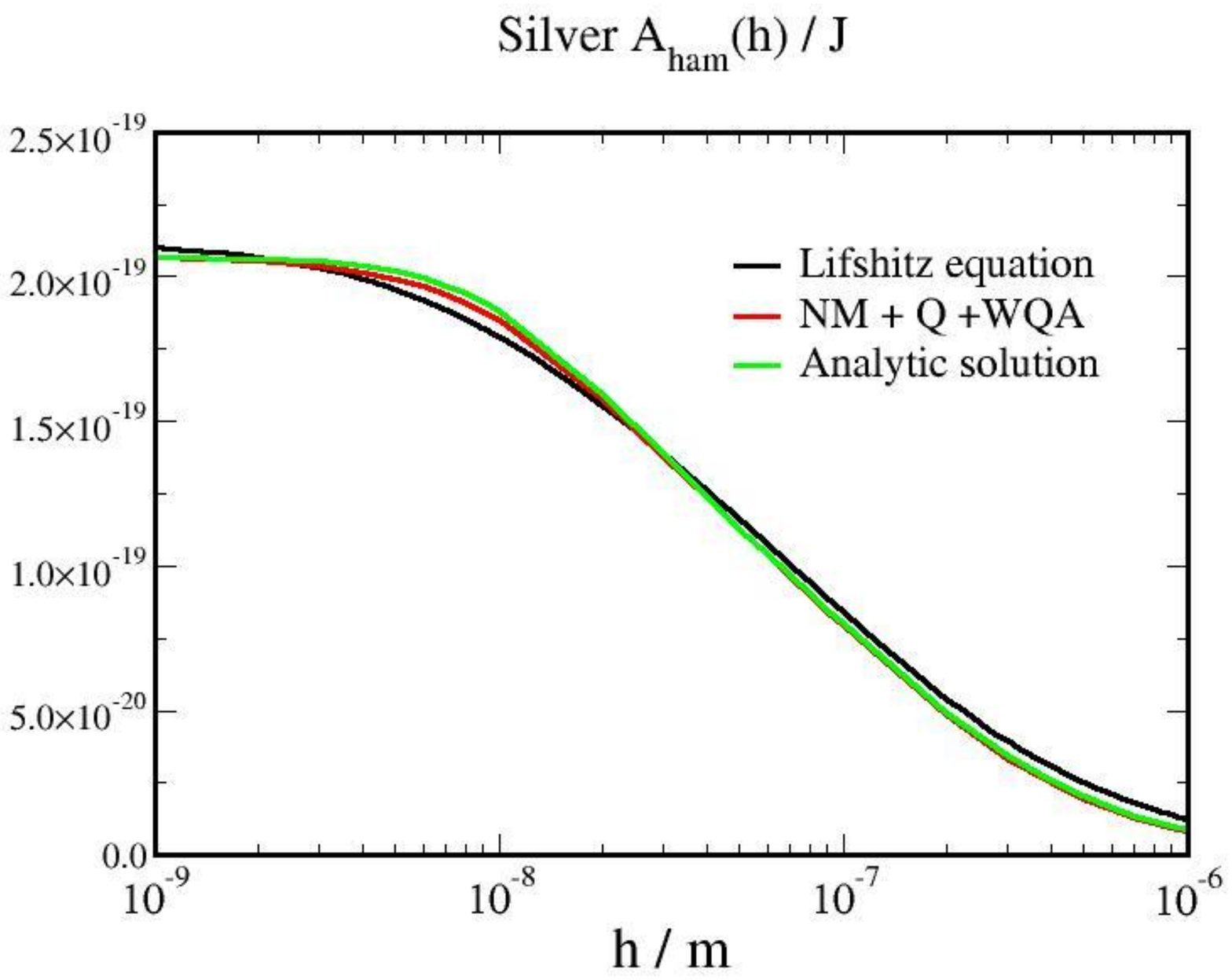}
	\includegraphics[width=0.48\textwidth,height=0.35\textwidth,keepaspectratio]{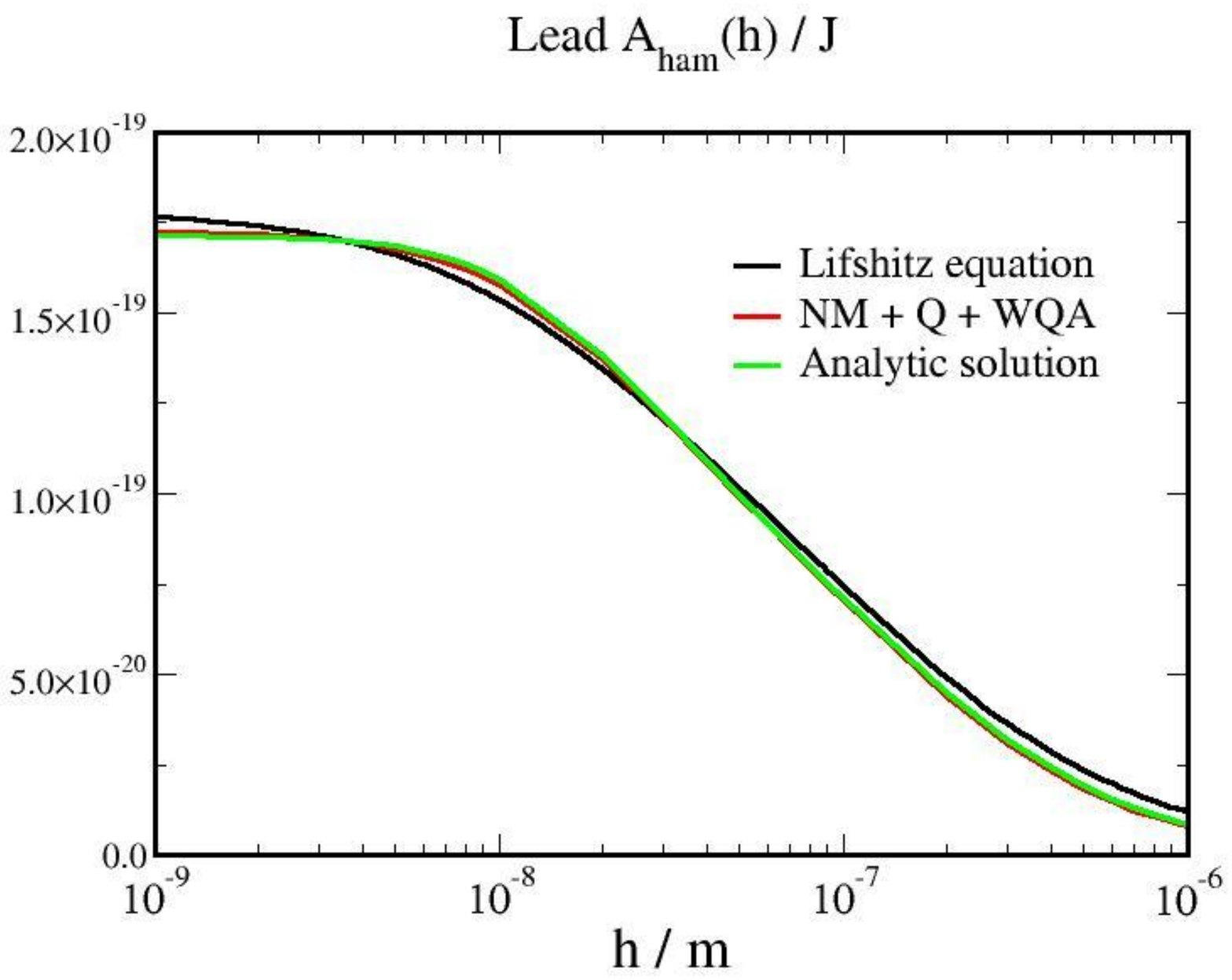}
	\\
	\includegraphics[width=0.48\textwidth,height=0.35\textwidth,keepaspectratio]{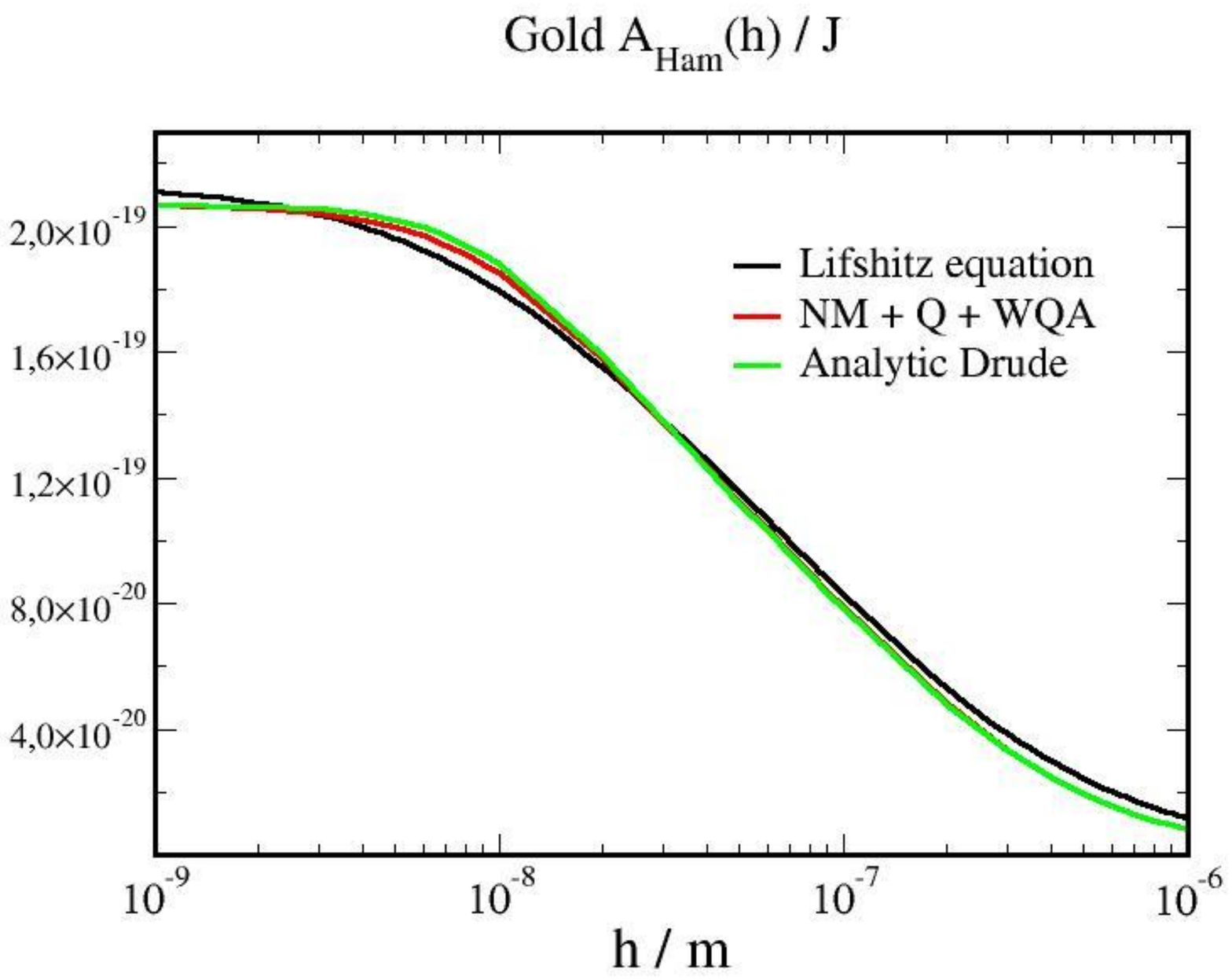}
	
	\caption{Retarded interaction coefficients computed with the Drude model
		at 300 K. Hamaker coefficients for the interaction of two plates of Al
		(top left), W (top right), Ag (bottom left) and Pb (bottom right), across
		vacuum. The $A_{ham}(h)$ are displayed in Joules. In each case, it is shown a comparison of the $A_{ham}(h)$ resulting from the Lifshitz equation (Eq. \ref{eq:lifshitz}) in black, the WQA with $\nu_{\infty}$ obtained via the Numerical Method to match the Quadrature $A_{ham,Q}(0)$ (NM + Q + WQA) in red, and the WQA supported by the analytic solution for $\nu_{\infty}$ in green.}
	\label{fig:drude}
\end{figure*}

\section{Analytic solution for the Drude model}

In the following, we will gauge Eq.\ref{eq:wqa} at $h=0$ with Eq.\ref{eq:quadrature} in order to estimate the system parameter $\nu_{\infty}$. Once this is known, we can then use Eq.\ref{eq:wqa} to estimate the Hamaker function for arbitrary values of the plate separation.

We begin the exploitation of the previous formulae by assuming the single - Drude model - oscillator as a sufficient description of the dielectric response of the metal. This model assumes free moving electrons\cite{youn07}, i.e. electrons displacing with no restoring force\cite{parsegian05}, oscillating at the plasma frequency, $\omega_{P}$. These electrons can experiment collisions with defects, lattice vibrations or with other electrons\cite{tanner13}, resulting in a damping coefficient, $\gamma$. All together, the Drude model reads
\begin{equation}
        \epsilon_{m}(i\omega) = 1 + \frac{\omega_{P}^{2}}{\gamma\omega+\omega^{2}}
        \label{eq:drudemodel}
\end{equation}
This representation has a fundamental limitation. It is not clear that the employment of the Drude model for real metals is physically justified, since most of them are  expected to exhibit  interband transitions\cite{ordal83}. This might be surpassed simply by considering that the Drude model is merely a phenomenological description of $\epsilon_{m}(i\omega)$, which is always a smooth function irrespective of the complex optical response of the metal.
\\
It is with such spirit that Ordal \textit{et al.} published\cite{ordal83} a set of Drude model parameters for various metals obtained from a fit of data in the near and far IR regime. These parameters are tabulated in Table \ref{table:drude} for Al, W, Ag and Pb, which also includes results from Gudarzi and Aboutalebi for Au \cite{gudarzi21}.
\\
However, it is clear that the Drude model does not properly account for the possible contributions of the high frequencies. This potentially has a great impact on the calculations surrounding the Lifshitz theory, because most of the Matsubara frequencies fall on a high energy regime. Even if the fact that metals are good conductors entails that the IR frequencies will strongly contribute to $A_{ham}(h)$, neglecting the remaining frequencies might provide a poor estimation. Hence the consequences of performing such approximation will be discussed later.
\vspace{0.1cm}
\\
Nevertheless, the use of the Drude model presents a very important advantage: it
allows us to get an explicit expression of $\nu_{\infty}$ for the interaction
between two plates of a certain metal across vacuum, depending fundamentally on
the plasma frequency of that metal\cite{supplementary21}. Indeed, assuming  $\omega_{P} \gg \gamma,\ \omega_{T}$, which is usually the case, Eq.\ref{eq:quadrature} provides the following simple result for the Hamaker constant between two metallic plates:
\begin{equation}
A_{ham}^{\omega_{n}>0}(h\to 0) =
\frac{3\, Li_3(5/8) }{5\sqrt{8}}   \,\hbar \omega_P
\label{eq:hamakerdrude}
\end{equation}

Matching this result to the $h\to 0$ limit of Eq.\ref{eq:wqa} provides a transcendental equation for $\nu_{\infty}$ which can be solved exactly to yield:
\begin{equation}
        \nu_{\infty} = f\nu_{P} - \nu_{T}
        \label{eq:drudeanalytic}
\end{equation}
Where $\nu_{P} = \frac{2}{c}\omega_{P}$ and $f = 0.553656$. Using Eq.\ref{eq:drudeanalytic} together with Eq.\ref{eq:wqa} provides a fully analytical description of the Hamaker function in all ranges of plate separation.
\\
We test this result for a number of metals. The analytical solution is compared with two numerical estimates of different level or refinement. Firstly, results are compared with the exact Hamaker function of Eq.\ref{eq:lifshitz}. Secondly, results are given for the WQA, Eq.\ref{eq:wqa}, with the parameter $\nu_{\infty}$ obtained by 
forcing $A_{ham}^{\omega > 0}(h\rightarrow 0)$ of the WQA (Eq.\ref{eq:wqa}) to match $A_{ham,Q}^{\omega_{n}\ >0}(h \rightarrow 0)$ of the Quadrature in Eq. \ref{eq:quadrature}. We call the latter prescription the NM + Q + WQA method. Table I displays values of $\nu_{\infty}$ that result from this prescription for a number of metals.

Fig. \ref{fig:drude} illustrates the results of this comparison.  The analytic solution shows excellent agreement with the two numerical methods for all tested metals, which highlights the accuracy of our closed form results Eq.\ref{eq:wqa} and Eq.\ref{eq:hamakerdrude}-Eq.\ref{eq:drudeanalytic}. 
\vspace{0.1cm}
\\
%\begin{table}[h!]
%\centering
%\begin{tabular}{|c|c|c|c|c|}
%\hline 
%	$ $ & $Al$ & $W$ & $Ag$ & $Pb$ \\ 
%\hline 
%	$\omega_{P}/eV$ & $14.78$ & $6.01$ & $9.01$ & $7.70$ \\ 
%\hline 
%$\gamma/eV$ & $8.04 \times 10^{-2}$ & $5.38\times10^{-2}$ & %$1.80\times 10^{-2}$ & $0.18$ \\
%\hline
%	$\nu_{\infty}/m^{-1}$ & $3.80\times 10^{7}$ & %$1.56\times 10^{7}$ & $2.32\times 10^{7}$ & $2.00\times %10^{7}$ \\
%\hline
%\end{tabular} 
%	\caption{Plasma frequency, $\omega_{P}$, and damping %coefficient, $\gamma$, of Al, W, Ag and Pb published by %Ordal \textit{et al}\cite{ordal83}. The last line shows the %$\nu_{\infty}$ parameter obtained with the Numerical %Method, employed to get the red lines in Fig. %\ref{fig:drude}.}
%\label{table:drude}
%\end{table}
%\vspace{0.1cm}
\\

\begin{table}[h!]
	\centering
	\begin{tabular}{|c|c|c|c|c|c|}
		\hline 
		$ $ & $Al$ & $W$ & $Ag$ & $Pb$ & $Au$\\ 
		\hline 
		$\omega_{P}/eV$ & $14.78$ & $6.01$ & $9.01$ & $7.70$  & $9.1$ \\ 
		\hline 
		$\gamma\cdot 10^2/eV$ & $8.04$ & $5.38$ & $1.80$ & $18$ & $6.0$ \\
		\hline
		$\nu_{\infty}\cdot 10^{-7}/m^{-1}$ & $3.80$ & $1.56$ & $2.32$ &
$2.00$ & $9.27$\\
		\hline
	\end{tabular} 
	\caption{Plasma frequency, $\omega_{P}$, and damping coefficient, $\gamma$ used in this work. Results for Al, W, Ag and Pb from Ref.\cite{ordal83}. Results for Au from Ref.\cite{gudarzi21}.The last line shows the $\nu_{\infty}$ parameter obtained with the Numerical Method, employed to get the red lines in Fig. \ref{fig:drude}.}
	\label{table:drude}
\end{table}

To check whether the Drude model is adequate enough to account for a complete description of $\epsilon_{m}(\omega)$, we take advantage of
improved dielectric parametrizations reported recently, which account for
interband transitions in the ultraviolet frequencies and
beyond.\cite{tolias18,gudarzi21} We use this
data to compute the retarded Hamaker coefficients with the Lifshitz equation,
and compare it with the output of the analytic solution for a single  Drude
oscillator.

\begin{figure*}[htb!]
\includegraphics[width=0.48\textwidth,height=0.35\textwidth,keepaspectratio]{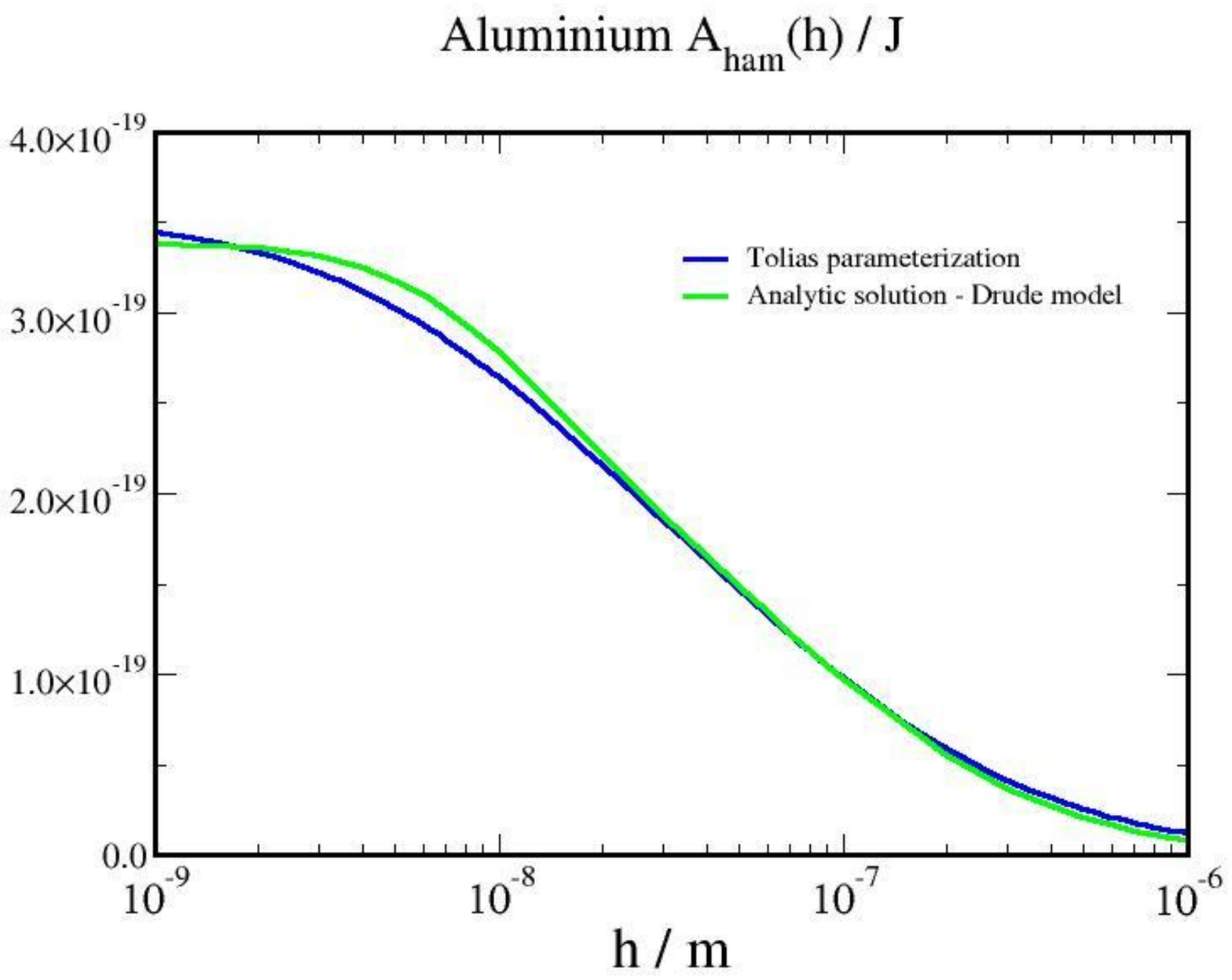}
\includegraphics[width=0.48\textwidth,height=0.35\textwidth,keepaspectratio]{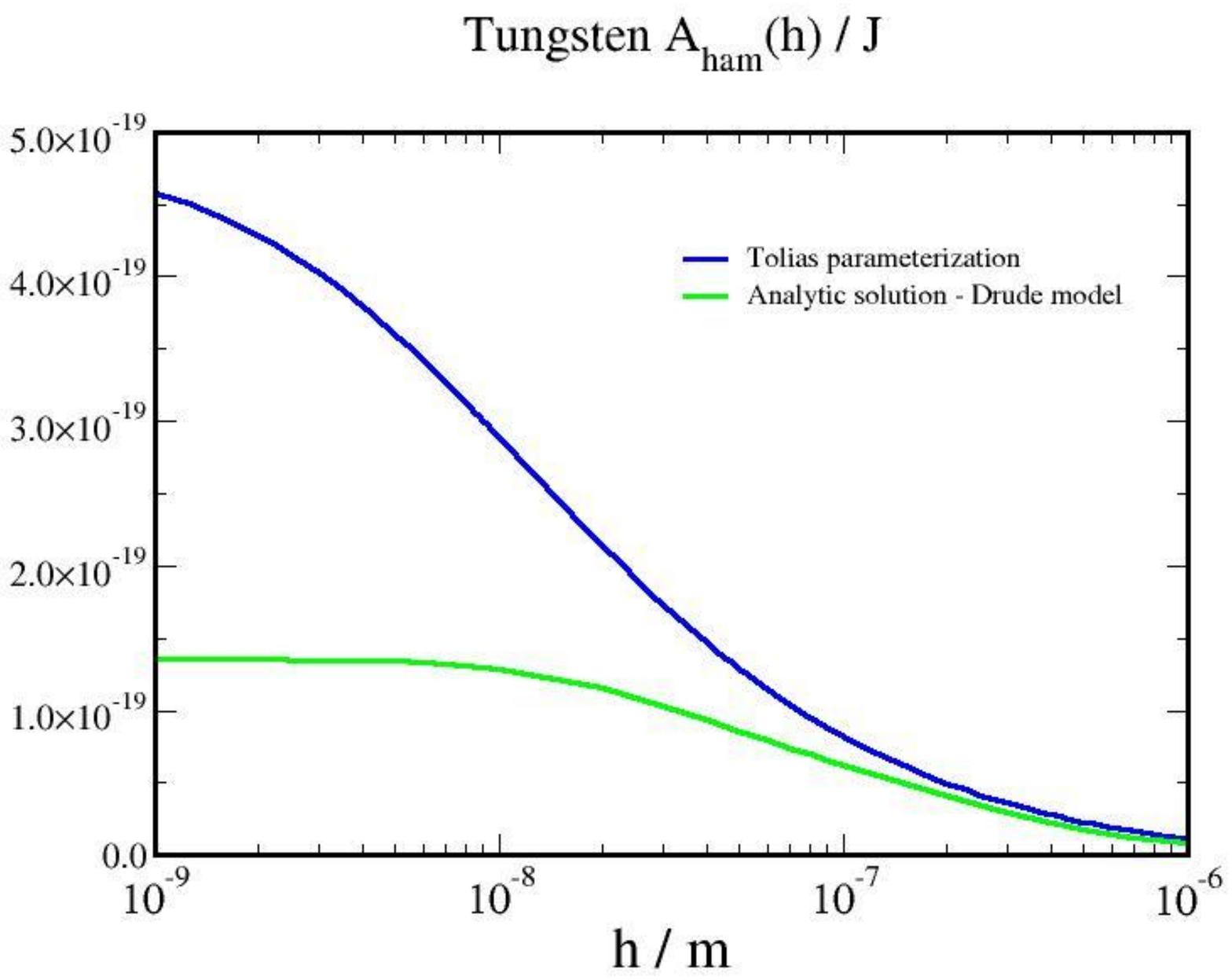}
	\caption{Retarded interaction coefficients of several metals
	   at 300 K.  The blue lines represent the Hamaker coefficients for the 
	   interaction
	of Al (left), and W (right), using the  Lifshitz equation and detailed
   dielectric data from Ref.~\cite{tolias18}. The analytic solution within the 
Drude model is also displayed in green for comparison. The Hamaker coefficients
are provided in Joule.}
\label{fig:tolias}
\end{figure*}

Fig.\ref{fig:tolias} displays $A_{ham}(h)$ of Al and W emerging from Eq. \ref{eq:lifshitz} with the parameterization of P. Tolias\cite{tolias18}, compared to the output of the analytic expression with the Drude model using the parameters published by Ordal \textit{et al}\cite{ordal83}. In the case of Al, the Drude model seems to provide a good characterization of $\epsilon_{m}(\omega)$, resulting in a very similar $A_{ham}(h)$ function for all separation distances. 
\vspace{0.1cm}
\\
On the contrary, for W, the high frequencies contributions to $A_{ham}(h)$ that are missed by the single Drude fit lead to a very low estimate of the Hamaker constant as  obtained from the analytic solution. As the retardation effect sets in, these large frequencies are cut off, and the remaining ones are properly given by the Drude model, so that the very large distances regime is well described with the single oscillator model of Ordal \textit{et al}. This highlights that the analytic solution will provide a poor approximation whenever the metal presents significant high frequency contributions to its dielectric response, revealing an insufficiency of the Drude model to account for those frequencies.

This observation is relevant for the experimental measurement of the Casimir regime of van der Waals forces.\cite{lamoreaux97,mohideen98,bressi02,munday09,man09,garret18} Particularly, high precision experiments aimed at testing the low temperature limit, Eq.2 often rely on the asymptotic expansion of Eq.3 based on a single Drude oscillator.\cite{} It is therefore very important to assess to what extent can one neglect contributions of interband transitions of small wave-length. We check this for the particularly significant case of gold, which is most often the choice for high precision measurements of the Casimir force.\cite{lamoreaux97,mohideen98,bressi02,munday09,man09,garret18} Fig.\ref{fig:gold} displays the exact Lifshitz result for gold, with dielectric properties as obtained in Ref.\cite{gudarzi21} The black line is the result obtained with the complete dielectric response, while the blue line provides the Hamaker function when only the conducting electrons are considered. At short distances, we see that more than half of the Hamaker constant results from contributions due to core electrons, as expected. Starting at about 100~nm, however, the contribution of core electrons has vanished almost completely. Whence, for gold it appears that asymptotic expansions based on simple Drude or plasma models should be reliable in the micrometer range. Care must be taken when asymptotic expansions are used below the hundreth of nanometer range, however.

\section{Comparison with numerical solutions}

So far we have tested the Quadrature proposed in 
Eq.~\ref{eq:quadrature} for metals described by
the Drude model, where the integrals implied by $I_{m}$ are analytically
solvable\cite{supplementary21}. 

As demonstrated in the preceding section, despite of its potential strength, the use of the analytic solution of the Drude model is quite circumstantial, depending on whether the high energy regime is negligible or not.

Now we show that the single-point quadrature of Eq.\ref{eq:quadrature} alone is
able to provide a simple method to get the Hamaker constant, $A_{ham}(0)$, upon
numerical integration, even for those cases where $\epsilon_{m}(\omega)$
exhibits a complex high-frequency behavior.  We use once again the fits for the
dielectric response of metals published by Tolias and Gudarzi and Aboutalebi to
get the $I_{1}$ and $I_{2}$ appearing in Eq. \ref{eq:quadrature} through
numerical integration with the composite trapezoidal
rule.\cite{tolias18,gudarzi21}
\vspace{0.1cm}
\\

\begin{figure*}[htb]
	\includegraphics[width=0.48\textwidth,height=0.35\textwidth,keepaspectratio]{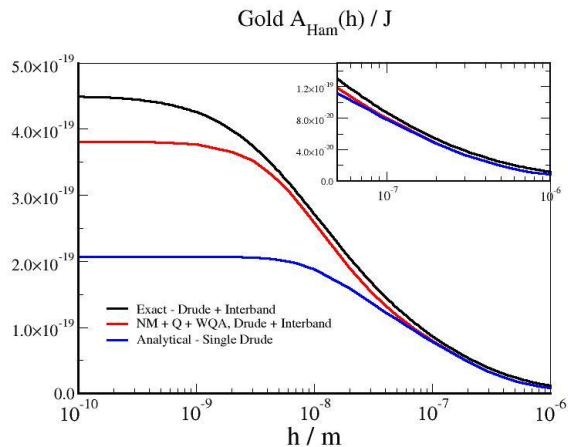}
	\caption{$A_{ham}(h)$ for the interaction of gold plates across air. The black line is obtained using Lifshitz theory, with full dielectric properties as reported in Ref \cite{gudarzi21}  (Drude oscillator + core electrons). The blue line are results obtained using only the Drude oscillator for conduction electrons. The red line is predictions from  NM + Q + WQA for the full dielectric response.
		\label{fig:gold}}
	\end{figure*}

\begin{figure*}[htb]
\includegraphics[width=0.48\textwidth,height=0.35\textwidth,keepaspectratio]{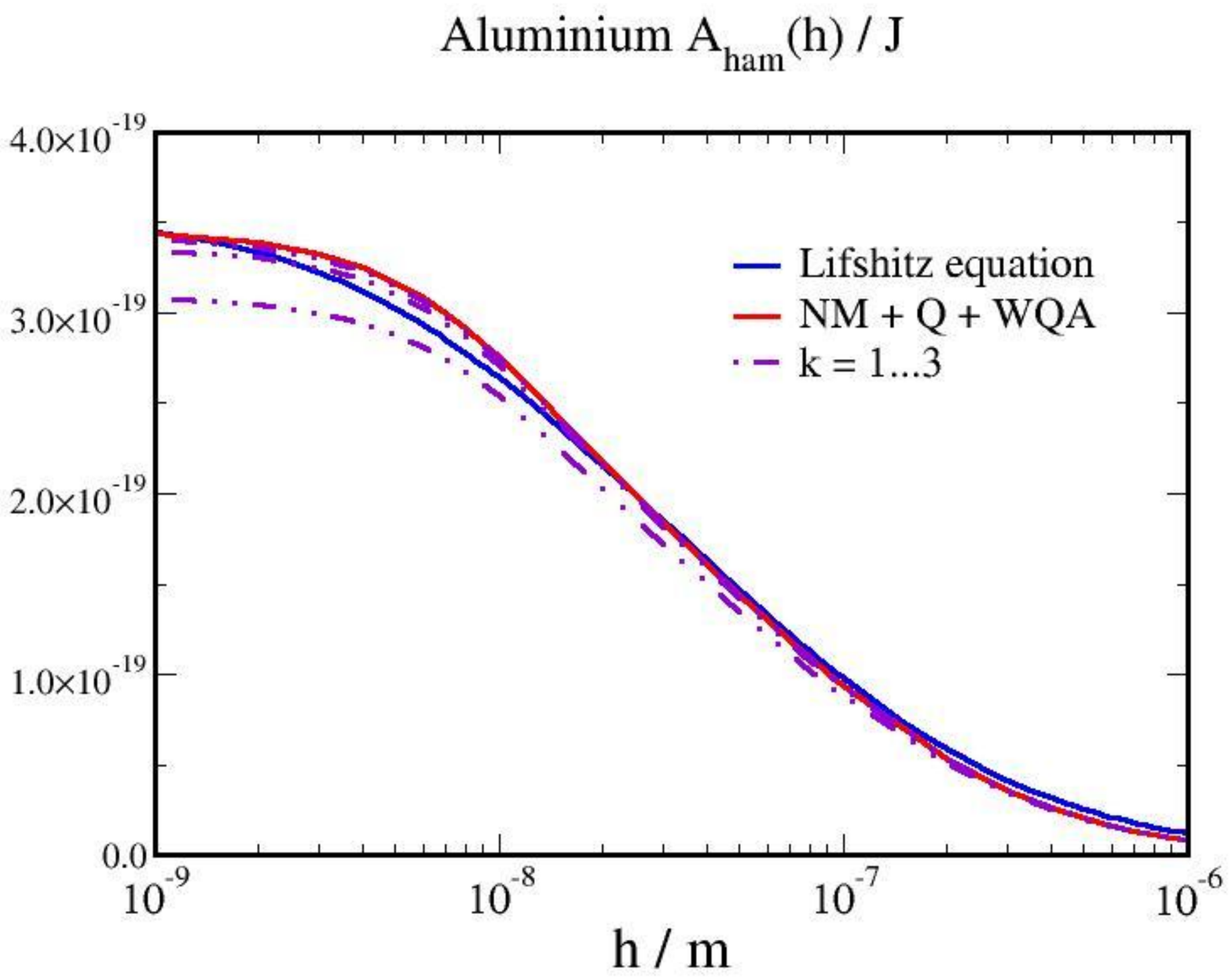}
\includegraphics[width=0.48\textwidth,height=0.35\textwidth,keepaspectratio]{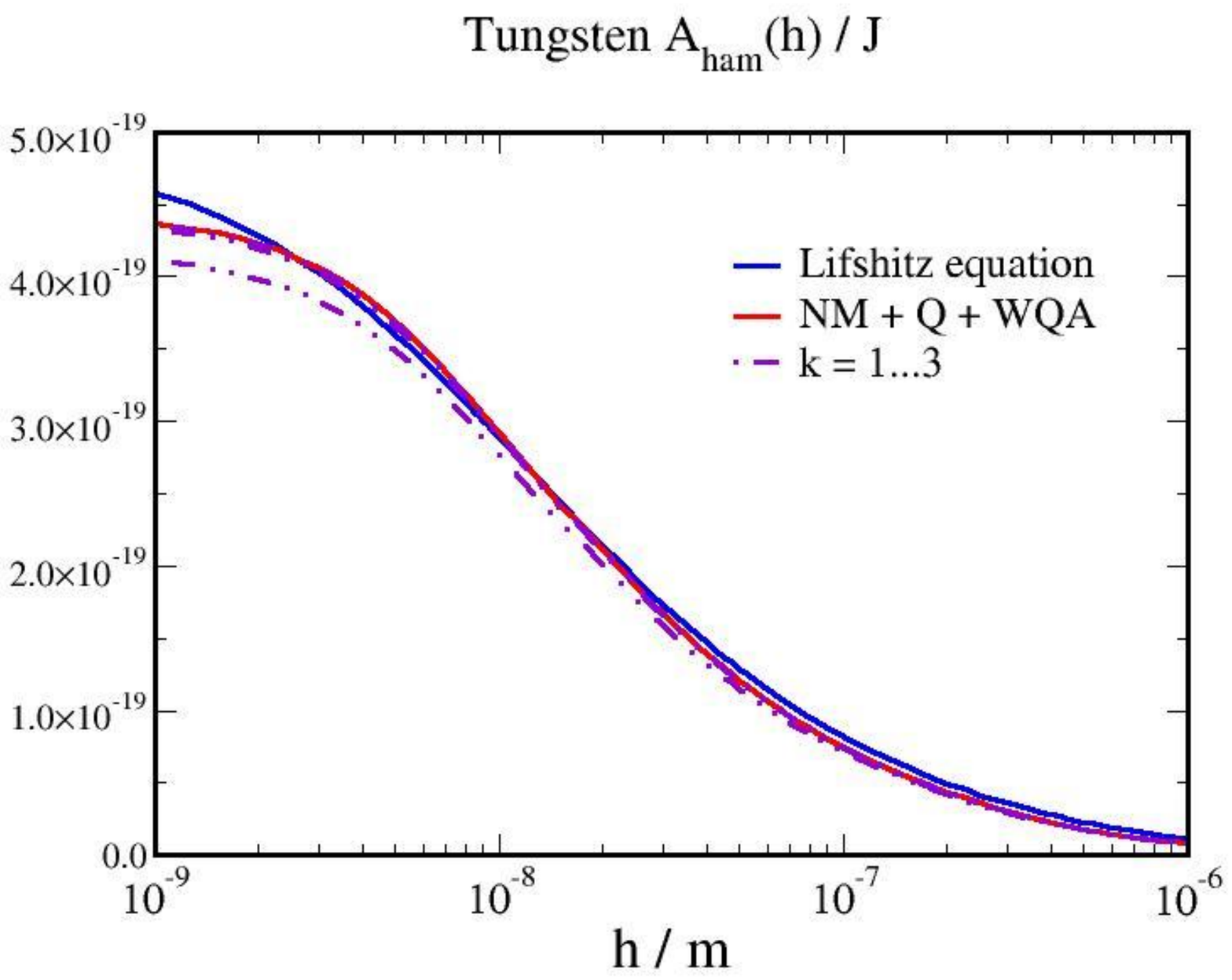}
	\caption{$A_{ham}(h)$ of Aluminium (left) and Tungsten (right) resulting
	from the Lifshitz formula (blue line) and the WQA (red line), with
   detailed optical properties as described in Ref.\cite{tolias18}. 
The $\nu_{\infty}$ parameter of the WQA has been obtained via Numerical Method
to match the Quadrature, and the dashed violet lines represent the WQA solution
with one, two, and three terms of the summation in k. Hamaker functions are
provided in Joule.}
	\label{fig:convergence}
\end{figure*}

\begin{table}[h!]
\centering
\begin{tabular}{|c|c|c|c|c|c|}
\hline  
	$ $ & $Al$ & $W$ & $Be$ & $Cr$ & $Au$\\ 
\hline 
	$A_{ham,Q}(0)/(10^{-19}\ J)$ & $3.44$ &  $4.74$ & $3.39$ & $3.61$ & $4.40$ \\
\hline
	$A_{ham}^{}(0)/(10^{-19} \ J)$ & $3.53$ & $4.85$ & $3.47$ & $3.69$ & $4.49$ \\
\hline
	r. e. $\%$ & $2.55$ & $2.27$ & $2.30$ & $2.17$ & $2.00$ \\
\hline
\end{tabular} 
	\caption{Hamaker constants for several metals as obtained from 
	   detailed optical properties of Ref.\cite{tolias18,gudarzi21}.
   The first line displays the Quadrature values computed with numerical
integration, where $A_{ham,Q}(0)=
A_{ham,Q}^{\omega_{n}>0}(0)+A_{ham}^{\omega_{n}=0}$. The second line shows exact
results from Lifshitz theory. Results for Al, W, Be, Cr from Ref.\cite{tolias18}. Results for Au calculated in this work using a dielectric functions as provided in Ref.\cite{gudarzi21}. The last line contains the relative error made by the Quadrature method.}
\label{table:tolias}
\end{table}

Fig.\ref{fig:convergence} represents the Hamaker function of Al, comparing the result of the exact Lifshitz equation with  the NM+Q+WQA prescription (Eq.\ref{eq:wqa} and Eq.\ref{eq:quadrature}). 
We plot in dashed purple lines the solution of Eq. \ref{eq:wqa} including an increasing number of terms in the summation over $k$ to check the convergence of the series. The second term provides already acceptable accuracy with respect to the complete summation, and almost complete convergence is achieved with the third term. This examination supports the use of a Quadrature exact up to second order, that was the assumption under which Eq. \ref{eq:quadrature} was derived. In fact,the difference between the second order ($k=2$) and third order results ($k=3$) is smaller than the uncertainty that results from the parameterization of the dielectric response in most systems\cite{burger20}.
\vspace{0.1cm}
\\
When numerically solving the equality between
$A_{ham,Q}^{\omega_{n}>0}(h\rightarrow 0)$ of the Quadrature and
$A_{ham}^{\omega_{n}>0}(h\rightarrow 0)$ of the WQA for tungsten, it was found
that there is no $\nu_{\infty}$ parameter that provides exactly that match.  In
this case, the validity of the Q+WQA method can be
compromised.  In practice, we find that taking the parameter that gives the larger possible $A_{ham}^{\omega_{n}>0}(h \rightarrow 0)$ already results in a very good approximation for both the value of the Hamaker constant and the habit of the Hamaker function in the Casimir regime. Indeed, the Fig. \ref{fig:convergence} shows that the WQA reproduces closely the functional behavior of the Hamaker function of the Lifshitz theory. We found that this problem does become quite significant for the relevant case of gold. Here, we did not find a choice of $\nu_{\infty}$ that could match the Hamaker constant, and the optimal value provides a Hamaker function that underestimates $A_{ham})(h\to 0)$ by 15\%, as shown in Fig.\ref{fig:gold}.
\vspace{0.15cm}
\\
Despite this defficiency in predicting the full Hamaker function for some metals, the proposed method for the calculation of Hamaker constants does remarkably well for all metals studied. Table \ref{table:tolias} displays the comparison between Hamaker constants of Al, W, Be, Cr and Au, and those obtained via Eq. \ref{eq:quadrature}. The Quadrature provides good agreement with the result of the Lifshitz equation without retardation, and its performance never exceeds a $3\%$ of relative error. This suggests the use of this novel Quadrature as a straightforwardly solvable alternative to the more intricate Lifshitz formula. The use of this formula is expected to be particularly helpful for estimating Hamaker constants between materials across a dielectric medium, where the first order approximation or Tabor-Winterton approximation often fails\cite{bergstrom97}.

\section{Conclusions}
\label{sec:conclusion}

Understanding the dispersive interactions between two surfaces is a crucial feature in the study of adhesion and friction phenomena\cite{israelachvili91,parsegian05,munday09}. At zero temperature, the van der Waals free energy exhibits a crossover from the non-retarded (London) behavior at short distances  to the retarded (Casimir) behavior at long distances of separation\cite{palasantzas08}. At finite temperature, however, the Casimir regime of retarded interactions is suppressed at sufficiently large distances.  This complex crossover behavior is non-trivially embodied in the Lifshitz equation for retarded interactions. The influence of the retardation effect switches the distance dependence of the interaction energy from the $\sim 1/h^{2}$ typical of London dispersion interaction, to the $\sim 1/h^{3}$ associated to the Casimir regime\cite{parsegian05} and back to$\sim 1/h^{2}$ at finite temperature and inverse distances smaller than a thermal wave-number $\nu_T$.
\vspace{0.1cm}
\\
In this paper, we have worked out an analytical approximation for the Hamaker function which illustrates the crossover behavior between these three different regimes and remains accurate at all ranges of separation. In this study, we also present an accurate Quadrature method to compute the Hamaker constant corresponding to the limit of small plate separation. Our quadrature rule consistently reproduces the values of $A_{ham}(0)$ provided by previous studies\cite{tolias18}, with less than 3\% error.
We have illustrated our results with the special case of two metallic plates in vacuum, but the method can be applied just as well for any two materials interacting across a dielectric medium.
\vspace{0.1cm}
\\
Finally, we made use of the Quadrature method to infer a fully analytical equation for the computation of the retarded interaction coefficients  between two metal plates with dielectric response as given  by a single Drude oscillator. However, we highlight that the use of this formula is limited to those metals with little response at large frequencies.
\vspace{0.1cm}
\\
We believe that this work provides a comprehensive picture of the behavior of the retarded interactions between metallic plates.
Methodologically, we hope that the quadrature rules employed here are susceptible of quantitative exploitation in a broad range of studies where dispersion interactions play an important role.
\section*{Acknowledgments}

We acknowledge funding from the Spanish Agencia Estatal de Investigaci\'on under grant FIS2017-89361-C3-2-P.

\section*{Authors Contributions}

JLM and LGM  discussed and formulated theory. JLM performed calculations and drafted manuscript. LGM designed research and revised manuscript.

%\bibliography{./referenc}% Produces the bibliography via BibTeX.


\clearpage

\newpage

\clearpage

\appendix

\onecolumngrid

\setcounter{page}{1}
\pagenumbering{arabic}

{\centering
   {\large Supporting Information for}

   {\Large Analytical theory for the crossover from retarded to non-retarded
	interactions between two metal plates 
   \\ by \\}
   {\large Juan Luengo-M\'arquez$^1$ and Luis G. MacDowell$^2$}

   {\normalsize
	$^1$Department of Theoretical Condensed Matter Physics and Instituto
	Nicol\'as Cabrera, \\
	Universidad Autónoma de Madrid, 28049 Madrid (Spain) 
	  \\
	  $^2$Departamento de Qu\'{\i}mica F\'{\i}sica, Facultad de Ciencias
   Qu\'{\i}micas, Universidad Complutense, Madrid, 28040, Spain.}

}

   \vspace*{1cm}

   %\begin{abstract}
   This document contains supporting information on the derivation of
   results from the main paper. To facilitate cross referencing, this
   materials is written as  an appendix section. The
   equation numbering and bibliography  follow the original paper, with
   equation labels and references
   not in this document  referring to those of the original paper.
   %\end{abstract}

\section{Lifshitz equation with retardation}
The Hamaker function of the exact Lifshitz equation for the interaction between two plates "$1$" and "$2$" composed of two arbitrary substances, across a medium "$med$" reads
\begin{equation}
   A_{ham}(h) = -6 h^{2} k_{B}T{\sum_{n=0}^{\infty}}'\int_{0}^{\infty}\rho \  d\rho \ln(D^{E}D^{M})
	\label{eq:exact_lifshitz}
\end{equation}
\begin{equation}
	D^{E,M} = 1 - \Delta_{1med}^{E,M}\Delta_{2med}^{E,M}e^{-2\rho_{med}h}
	\notag
\end{equation}
\begin{equation}
	\Delta_{imed}^{E} = \frac{\rho_{med}-\rho_{i}}{\rho_{med}+\rho_{i}}, \hspace{0.75cm} \Delta_{imed}^{M} = \frac{\epsilon_{i}\rho_{med}-\epsilon_{med}\rho_{i}}{\epsilon_{i}\rho_{med}+\epsilon_{med}\rho_{i}} 
	\notag
\end{equation}
\begin{equation}
	\rho_{i,med}^{2} = \rho^{2} + \frac{\epsilon_{i,med}\omega_{n}^{2}}{c^{2}}, \hspace{0.75cm} \omega_{n} = \frac{2\pi k_{B}T}{\hbar}n
	\notag
\end{equation}
Being $T$ the temperature in Kelvin, $k_{B}$ the Boltzmann constant, $c$ the speed of light, $\hbar$ the Planck constant in units of angular frequency, and $h$ the separation between the two plates, i. e. the thickness of the medium in between. Where the prime after the summation means that the $n=0$ term has an additional factor of 1/2.  The dielectric responses, $\epsilon_{i,med}$, are evaluated at the discrete Matsubara frequencies, $\omega_{n}$. Finally, $D^{E,M}$ expresses a condition imposed by the geometry of the system that every electromagnetic wave crossing through it must fulfill. When equalized to zero, it is dubbed the dispersion relation of the system.
\vspace{0.15cm}
\\
After splitting $\ln(D^{E}D^{M})=\ln(D^{E}) + \ln(D^{M})$ in Eq. \ref{eq:exact_lifshitz} and performing a Taylor expansion on each logarithm, we readily get
\begin{equation}
   A_{ham}(h) = 6 h^{2} k_{B}T{\sum_{n=0}^{\infty}}'\int_{0}^{\infty}\rho \  d\rho \sum_{k=1}^{\infty}\frac{(\Delta_{1med}^{E}\Delta_{2med}^{E})^{k}+(\Delta_{1med}^{M}\Delta_{2med}^{M})^{k}}{k}e^{-2k\rho_{med}h}
        \label{eq:expansion_lifshitz}
\end{equation}
We proceed now by changing the variable of the integral twice, first through $\rho\rightarrow \rho_{med}$, and then through $2\rho_{med}h \rightarrow x$. This procedure yields
\begin{equation}
   A_{ham}(h) = \frac{3 k_{B}T}{2}{\sum_{n=0}^{\infty}}'\int_{r_{n}}^{\infty}\sum_{k=1}^{\infty}x \ dx\frac{(\Delta_{1med}^{E}\Delta_{2med}^{E})^{k}+(\Delta_{1med}^{M}\Delta_{2med}^{M})^{k}}{k}e^{-k x}
        \label{eq:main_lifshitz}
\end{equation}
\begin{equation}
	r_{n} = \frac{2h\epsilon_{med}^{1/2}\omega_{n}}{c}
	\notag
\end{equation}
\begin{equation}
	\Delta_{imed}^{E} = \frac{x_{med}-x_{i}}{x_{med}+x_{i}}, \hspace{0.75cm} \Delta_{imed}^{M} = \frac{\epsilon_{i}x_{med}-\epsilon_{med}x_{i}}{\epsilon_{i}x_{med}+\epsilon_{med}x_{i}} 
	\notag
\end{equation}
\begin{equation}
	x_{i}^{2} = x^{2} + (\epsilon_{i}-\epsilon_{med})\left(\frac{2 h \omega_{n}}{c}\right)^{2}
	\notag
\end{equation}
Note that Eq. 3 and 4 of the main text are recovered after particularizing for the specific system under consideration with $\epsilon_{med} = \epsilon_{v}$, $x_{med}=x_{v}$, $\epsilon_{1}=\epsilon_{2}=\epsilon_{m}$, and $x_{1} = x_{2} = x_{m}$. All along this Supplementary Material we will address the general notation employed so far to extend certain results to differently labelled geometries, but the particularization is straightforwardly derived after substitution of the previous equalities. 
\vspace{0.15cm}
\\
The Eq. 5 of the main text is achieved by separating the $n=0$ term of Eq. \ref{eq:main_lifshitz}. Then $x_{i,med}=x$, so that all $\Delta_{imed}^{E}$ vanish, and the term with $\Delta_{imed}^{M}$ becomes constant in $x$. The resulting integral is easily solvable, and we reach
\begin{equation}
	A_{ham}^{\omega_{n}=0} = \frac{3k_{B}T}{4}\sum_{k=1}^{\infty}\left(\frac{\epsilon_{1}(0)-\epsilon_{med}(0)}{\epsilon_{1}(0)+\epsilon_{med}(0)}\right)^{k}\left(\frac{\epsilon_{2}(0)-\epsilon_{med}(0)}{\epsilon_{2}(0)+\epsilon_{med}(0)}\right)^{k}\frac{1}{k^{3}}
	\label{eq:static}
\end{equation}
Where only the static dielectric responses appear, i. e. the responses at zero frequency.
\section{Weighted Quadrature Approximation}
The WQA takes advantage of the one point Gaussian Quadrature used as an
analytical tool to compute the integrals in Eq. \ref{eq:main_lifshitz}.
Generally, as a numerical tool,  the N points Gaussian Quadrature 
is written as:
\begin{equation}
\int_{a}^{b}f(x)w(x)dx = \sum_{i=1}^{N}f(x_{i})m_{i}
\label{eq:npoints_gaussian}
\end{equation}
Where the smooth function $f(x)$ is evaluated at the nodes $x_{i}$, and $w(x)$
is the weight function of the quadrature. The set of nodes and weights $x_i$,
$m_i$ are chosen such that the integrals
\begin{equation}
I_{j} = \int_{a}^{b}x^{j}w(x)dx 
%= \sum_{i=1}^{N}x_{i}^{j}m_{i}
\notag
\end{equation}
are exact up to order $2N-1$.
Then the one point Gaussian Quadrature simply yields
\begin{equation}
\int_{a}^{b}f(x)w(x)dx = f(x_{1})m_{1}
\label{eq:onepoint_gaussian}
\end{equation}
\begin{equation}
	I_{0} = \int_{a}^{b}w(x)dx, \hspace{0.75cm}  I_{1} = \int_{a}^{b}x w(x)dx
\notag
\end{equation}
\begin{equation}
	m_{1} = I_{0}, \hspace{0.75cm}  x_{1} = \frac{I_{1}}{I_{0}}
\notag
\end{equation}
We take the $\omega_{n}>0$ contributions to $A_{ham}(h)$ in Eq. \ref{eq:main_lifshitz}, and then define $w(x)=x e^{-kx}$ as the weight function. Thus we have that
\begin{equation}
	I_{0} = \int_{r_{n}}^{\infty} x e^{-kx}dx = \frac{(k r_{n} + 1)}{k^{2}}e^{-k r_{n}}
	\label{eq:fgq}
\end{equation}
\begin{equation}
	I_{1} = \int_{r_{n}}^{\infty} x^{2} e^{-k x} dx = \frac{(k^{2} r_{n}^{2} + 2kr_{n} + 2)}{k^{3}}e^{-kr_{n}}
\notag
\end{equation}
From which we get the corresponding $m_{1}$ and $x_{1}$. The integral in $x$ in Eq. \ref{eq:main_lifshitz} is computed through the quadrature, and after extracting the summation in $k$ we reach
\begin{equation}
	A_{ham}^{\omega_{n}>0}(h) = \frac{3 k_{B}T}{2}\sum_{k=1}^{\infty}\sum_{n=1}^{\infty}R_{k}(n,x_{1,k},h)[k r_{n} + 1]e^{-kr_{n}}
        \label{eq:fqa_lifshitz}
\end{equation}
\begin{equation}
	R_{k}(n,x,h) = \frac{(\Delta_{1med}^{E}\Delta_{2med}^{E})^{k}+(\Delta_{1med}^{M}\Delta_{2med}^{M})^{k}}{k^{3}}
        \notag
\end{equation}
\begin{equation}
	x_{1,k} = \frac{kr_{n}^{2}+2r_{n}+2/k}{kr_{n}+1}
        \notag
\end{equation}
Recall that $\Delta^{E,M}_{imed} = \Delta^{E,M}_{imed}(n,x)$. Next we aim to approximate the summation in $n$ to an integral, for what we compute first the first-order correction of the Euler-MacLaurin formula
\begin{equation}
F = \sum_{n=a}^{b}f(n) = \int_{a}^{b}f(n)dn + \frac{1}{2}(f(a)+f(b))+\sum^{\infty}_{k=1} \frac{B_{2k}}{2k!}(f^{(2k-1)}(b)-f^{(2k-1)}(a))
\label{eq:euler_maclaurin}
\end{equation}
\begin{equation}
	\Delta F_{1st-order} = \frac{1}{2}(f(a)+f(b))+ \frac{1}{12}(f'(b)-f'(a))
\notag
\end{equation}
Where $B_{i}$ are the Bernoulli coefficients and $f^{(2k-1)}(n)$ represents the $(2k-1)$th derivative of $f(n)$. Note that the algebraic evolution of $R_{k}(n,x,h)$ in $n$ can be taken as constant compared to the exponential decay of $(kr_{n} -1)e^{-kr_{n}}$. Under this consideration, the computation of the first-order correction term of the Euler-MacLaurin formula for $f(n)=(kr_{n} -1)e^{-kr_{n}}$ is just
\begin{equation}
	\Delta F_{1st-order}(h) = \left[\frac{(kr_{T}+1)}{2}+\frac{(kr_{T})^{2}}{12}\right]e^{-kr_{T}} 
	\label{eq:correction_term}
\end{equation}
Whose dependence on $h$ comes by the hand of $r_{T}(h) = r_{n=1}(h)$. We highlight here such dependence because for most distances $r_{T} \ll 1$, and $\Delta F_{1st-order}$ is negligible compared to the leading term. The correction term only becomes relevant in the long distances regime, where $A_{ham}^{\omega_{n}=0}$ provides the dominant contribution of the van der Waals interaction, so we can neglect $\Delta F_{1st-order}$ safely and approximate
\begin{equation}
        A_{ham}^{\omega_{n}>0}(h) \approx \frac{3 k_{B}T}{2}\sum_{k=1}^{\infty}\int_{n=1}^{\infty}R_{k}(n,x_{1,k},h)[k r_{n} + 1]e^{-kr_{n}}dn
        \label{eq:fgq_lifshitz_integral}
\end{equation}
Now we change the variable as $r_{n} \rightarrow \nu h$, so that $\nu(n) = (4\pi k_{B}T\epsilon_{med}^{1/2}n)/(c\hbar)$, and we define $\nu_{T}=\nu(n=1)$. As $\epsilon_{med}$ is in general also a function of $n$, this transformation implies that
\begin{equation}
d\nu = \frac{4\pi k_{B}T}{c\hbar}\epsilon^{1/2}_{med}dn\left[1 + \frac{1}{2}\frac{d\ln\epsilon_{med}}{d\ln\omega_{n}}\right]
\label{eq:change_variable}
\end{equation}
And we name the term inside the brackets as $j_{med}$. Plugging this in Eq. \ref{eq:fgq_lifshitz_integral} and defining $\widetilde{R}_{k}(\nu,x,h) = \epsilon_{med}^{-1/2}j_{med}^{-1}R_{k}(\nu,x,h)$ leads to
\begin{equation}
	A_{ham}^{\omega_{n}>0}(h) = \frac{3 c\hbar}{8\pi}\sum_{k=1}^{\infty}\int_{\nu_{T}}^{\infty}\widetilde{R}_{k}(\nu,x_{1,k},h)[k\nu h + 1]e^{-k\nu h}d\nu
        \label{eq:fgq_lifshitz_changevar}
\end{equation}
The particularization along the same line described above recovers the Eq. 6 of the main text. It is in this stage that we introduce the material parameter $\nu_{\infty}$ via the inclusion of the factor $e^{\nu/\nu_{\infty}}e^{-\nu/\nu_{\infty}}$ inside the integrand. This allows the evaluation of the integral through a Gaussian Quadrature with the weight function $w(\nu) = e^{-\nu/\nu_{\infty}}[k\nu h+1]e^{-k\nu h}$, that drops exponentially as $\nu$ approaches $\nu_{\infty}$. Thus we solve
\begin{equation}
	I_{0} = \int_{\nu_{T}}^{\infty}e^{-\nu/\nu_{\infty}}[k\nu h +1]e^{-k\nu h}d\nu =
        \label{eq:sgq}
\end{equation}
\begin{equation}
 = \nu_{\infty}\frac{(\nu_{T}kh+1)(\nu_{\infty}kh+1)+\nu_{\infty}kh}{(\nu_{\infty}kh+1)^{2}}e^{-\nu_{T}kh-\frac{\nu_{T}}{\nu_{\infty}}}
\notag
\end{equation}

\begin{equation}
	I_{1} = \int_{\nu_{T}}^{\infty}\nu e^{-\nu/\nu_{\infty}}[k\nu h+1]e^{-k\nu h}d\nu =
\notag
\end{equation}
\begin{equation}
	 = \nu_{\infty}\frac{(\nu_{T}kh+1)(\nu_{\infty}kh+1)^{2}\nu_{T}+  (2\nu_{T}kh+1)(\nu_{\infty}kh+1)\nu_{\infty}  + 2\nu_{\infty}^{2}kh}{(\nu_{\infty}kh+1)^{3}}e^{-\nu_{T}kh-\frac{\nu_{T}}{\nu_{\infty}}}
\notag
\end{equation}
We name the quadrature point $\nu_{k}^{*} = I_{1}/I_{0}$ to avoid any confusion with $\nu_{T}$. Besides, we define $\xi_{k} = (\nu_{k}^{*}-\nu_{T})/\nu_{\infty}$ for compactness in the formula. The resulting expression is readily particularized to get the Eq. 8 of the main text.

\section{Quadrature for the Hamaker constant}
The derivation of the Quadrature expression for the Hamaker constant departs from Eq. \ref{eq:fgq_lifshitz_changevar} in the $h\rightarrow 0$ limit. We further change the variable as $\nu \rightarrow 2\epsilon_{med}^{1/2}\omega/c$, with $\omega_{T}=\omega_{n=1}$, and introduce again the summation inside the integral, to reach
\begin{equation}
	A_{ham}^{\omega_{n}>0}(h\rightarrow 0) = \frac{3 \hbar}{4\pi}\int_{\omega_{T}}^{\infty}\sum_{k=1}^{\infty}R_{k}(\omega,h\rightarrow 0)d\nu
        \label{eq:departure_quadrature}
\end{equation}
\begin{equation}
	R_{k}(\omega,h\rightarrow 0) =\left[\left(\frac{\epsilon_{1}-\epsilon_{med}}{\epsilon_{1}+\epsilon_{med}}\right)\left(\frac{\epsilon_{2}-\epsilon_{med}}{\epsilon_{2}+\epsilon_{med}}\right)\right]^{k} \frac{1}{k^{3}}
	\notag
\end{equation}
Note that, as explained in the main text, $R_{k}(\omega,x,h\rightarrow 0)$ no longer depends on $x$. Also, recall that the dependence on $\omega$ enters by the hand of $\epsilon_{i,med}=\epsilon_{i,med}(\omega)$.
\vspace{0.15cm}
\\
The first mean value theorem states that, given the integral of a certain function, $f(x)$, times a probability density function, $w(x)$, there is a certain point, say $x^{*}$, such that
\begin{equation}
	\int_{a}^{b}f(x)w(x)dx = f(x^{*})\int_{a}^{b}w(x)dx
	\label{eq:mvt}
\end{equation}
This generalized form of the first mean value theorem holds as long as $f(x)$ is a continuous real-valued function and $w(x)$ is an integrable nonnegative function. Moreover, realize that if $w(x)$ is the PDF of the continuous uniform distribution, then the usual form of the first mean value theorem is recovered. Also, we define $f(x^{*})\equiv f^{*}$ henceforth.
\vspace{0.15cm}
\\
We apply this version of the first mean value theorem to Eq. \ref{eq:departure_quadrature}. We begin developing the integrand up to second order of the summation as
\begin{equation}
	\sum_{k=1}^{\infty}R_{k}(\omega,h\rightarrow 0) = \left(\sum_{k=1}^{\infty}\frac{R_{1}(\omega)^{k-1}}{k^{3}}\right)R_{1}(\omega) 
        \label{eq:transformation_integrand}
\end{equation}
Where $R_{1}(\omega)=R_{k=1}(\omega)$. 
We proceed by employing the first mean value theorem with $w(x)\leftrightarrow
R_{1}(\omega)$, and requiring the output to match the exact solution of the
original integral up to second order in the expansion. This allows us to write $f^{*}\leftrightarrow R^{*}$ in terms of the resulting integrals, and then extend it to the complete series in Eq. \ref{eq:transformation_integrand}. This way, the Eq. 12 of the main text is achieved, that is exact up to second order by construction, but approximates very well the rest of the series. The Hamaker constant of the Quadrature is consequently given in terms of the trilogarithm function of $R^{*}$
\begin{equation}
	Li_{3}(x) = \sum_{k=1}^{\infty}\frac{x^{k}}{k^{3}}
	\label{eq:trilogarithm}
\end{equation}

\section{Analytic solution within the Drude model}

Here we particularize for the system that we handle by stating that the medium in between the two plates is vacuum, and that both plates are composed of the same metal. Consequently, $\epsilon_{1,2}(\omega) = \epsilon_{m}$, $\epsilon_{med}(\omega) = \epsilon_{v}(\omega) = 1.0$, and $j_{med} = j_{v} = 1.0$. Within the Drude model, $R_{k=1}(\omega,h\rightarrow 0)$ in Eq. \ref{eq:departure_quadrature} becomes
\begin{equation}
	R_{1}(\omega,h\rightarrow 0) = \left(\frac{\omega_{P}^{2}}{\omega_{P}^{2}+2\gamma\omega + 2\omega^{2}}\right)^{2}
	\label{eq:erre_drude}
\end{equation}
Where $\omega_{P}$ and $\gamma$ are the plasma frequency and the damping coefficient, respectively. As explained in the main text, these are the parameters of the Drude model that characterize the dielectric response of the specific metallic species of the plates. The previous can be further developed approximating that for low $\omega$, $\omega_{P}^{2} \gg 2\gamma\omega$, and for large $\omega$, $2\omega^{2} \gg 2\gamma\omega$. Then
\begin{equation}
	R_{1}(\omega,h\rightarrow 0) \approx \left(\frac{\omega_{P}^{2}}{\omega_{P}^{2}+2\omega^{2}}\right)^{2} = \left(\frac{1}{1+2(\omega/\omega_{P})^{2}}\right)^{2} 
	\notag
\end{equation}
In the following lines, we demonstrate that an analytic expression of $\nu_{\infty}$ for the Drude model can be achieved by requiring the Hamaker constant of the WQA to match the Quadrature of the previous section. First, we take the $h\rightarrow 0$ limit of the WQA in Eq. 8 of the main text. 
\vspace{0.15cm}
\\
We write $R_{k}(\omega,h\rightarrow 0)$ in terms of the preceding result for $k=1$. In the WQA, this function is evaluated at $(\omega_{T}+\omega_{\infty})=(\nu_{T}+\nu_{\infty})\frac{c}{2}$. Subsequently, we define $f=(\omega_{T}+\omega_{\infty})/\omega_{P}$ to reach
\begin{equation}
	A_{WQA}^{\omega_{n}>0}(h\rightarrow 0) \approx \frac{3e\hbar}{4\pi}f\omega_{P}\sum_{k=1}^{\infty}\frac{1}{k^{3}}\left(\frac{1}{1+2f^{2}}\right)^{2k}
	\label{eq:wqa_limit}
\end{equation}
Where $e$ is the Euler's number and $\omega_{\infty}\approx f\omega_{P}$ has been assumed, neglecting $\omega_{T}$ for the sake of simplicity. The next step is to get the corresponding Hamaker constant of the Quadrature based on the first mean value theorem, for what we realize that $I_{1}$ and $I_{2}$ in Eq. 15 of the main text are analytically solvable for the $R_{1}(\omega,h\rightarrow 0)$ provided by Eq. \ref{eq:erre_drude}, namely
\begin{equation}
	I_{1} = \frac{\omega_{P}^{4}}{\Delta}\left[\frac{4\pi}{\Delta^{1/2}}-\frac{g'(\omega_{T})}{g(\omega_{T})}-\frac{8}{\Delta^{1/2}}\arctan{\left(\frac{g'(\omega_{T})}{\Delta^{1/2}}\right)}\right]
	\label{eq:analytic_integrals}
\end{equation}
\begin{equation}
	I_{2} = \frac{\omega_{P}^{8}}{3\Delta^{3}}\left[\frac{480\pi}{\Delta^{1/2}} - \frac{960}{\Delta^{1/2}}\arctan{\left(\frac{g'(\omega_{T})}{\Delta^{1/2}}\right)}-\frac{120 g'(\omega_{T})}{g(\omega_{T})}-\frac{10\Delta g'(\omega_{T})}{g^{2}(\omega_{T})} - \frac{\Delta^{2}g'(\omega_{T})}{g^{3}(\omega_{T})}\right]
	\notag
\end{equation}
\begin{equation}
	\Delta = 8\omega_{P}^{2}-4\gamma^{2},\hspace{0.5cm} g(\omega_{T}) = 2\omega_{T}^{2}+2\gamma\omega_{T}+\omega_{P}^{2}, \hspace{0.5cm} g'(\omega_{T}) = 4\omega_{T}+2\gamma
	\notag
\end{equation}
These solutions allows us to get the Hamaker constant of the Quadrature exactly, but aimed at providing an analytic formula for $\nu_{\infty}$, we must approximate $\omega_{P}^{2} \gg \gamma^{2}, \ \gamma\omega_{T}, \ \omega_{T}^{2}$. Under this condition, the previous integrals get simplified, and $R^{*}$ in Eq. 15 of the main text is easily evaluated. 

The purpose of the approximations that we have carried out is now unveiled. When we require $A_{WQA}^{\omega_{n}>0}(h\rightarrow 0)$ of Eq. \ref{eq:wqa_limit} to match $A_{ham}^{\omega_{n}>0}(h\rightarrow 0)$ of Eq. 17 of the main text, the resulting equality is a transcendental equation for $f$ with a few dimensionless factors
\begin{equation}
	f\sum_{k=1}^{\infty}\frac{1}{k^{3}}\left(\frac{1}{1+2f^{2}} \right)^{2k} = \frac{\sqrt{2}\pi}{5e}Li_{3}(5/8)
	\label{eq:transcendental_f}
\end{equation}
We solve Eq. \ref{eq:transcendental_f} numerically to get two possible solutions. Either $f_{1}=0.250141$ or $f_{2}=0.553656$ fulfill the condition imposed by Eq. \ref{eq:transcendental_f}. This happens because $\nu_{\infty}R_{k}(\nu_{T}+\nu_{\infty})$ is a non-monotonic function of $\nu_{\infty}$, but only the $\nu_{\infty}$ that emerges from $f_{2}$ preserves the desired habit of $A_{ham}^{\omega_{n}>0}(h)$.
\vspace{0.15cm}
\\
Finally, we return to the definition of $f$ to achieve
\begin{equation}
	\omega_{\infty} = f\omega_{P} - \omega_{T}
\label{eq:analytic_drude}
\end{equation}
With $f=0.553656$, which is the Eq. 18 of the main text. The accuracy of the approximations performed is verified in Fig. 1 of the main text.
\end{document}